\documentclass[10pt]{article}
\usepackage{cite}
\usepackage{amsmath,amssymb,amsfonts}
\usepackage{algorithmic}
\usepackage{graphicx}
\usepackage{textcomp}
\usepackage{upgreek}
\usepackage{mathabx}
\usepackage{amsbsy}
\usepackage{color}
\usepackage{multirow}

\newtheorem{Theorem}{Theorem}

\newtheorem{Assumption}{Assumption}
\newtheorem{Lemma}{Lemma}

\newtheorem{Corollary}{Corollary}
\newtheorem{Remark}{Remark}

\def\Nat{\mathbb{N}}
\def\Real{\mathbb{R}}
\def\Comp{\mathbb{C}}
\def\P{\mathbb{P}}
\def\E{\mathbb{E}}
\def\Ac{\mathcal{A}}
\def\Dc{\mathcal{D}}
\def\Ec{\mathcal{E}}
\def\Fc{\mathcal{F}}
\def\Gc{\mathcal{G}}
\def\Lc{\mathcal L}
\def\Tc{\mathcal T}
\def\Lcb{\bar{\mathcal L}}

\def\Nc{\mathcal{N}}
\def\Ncb{\overline{\mathcal{N}}}

\def\Sc{\mathcal S}
\def\Vc{\mathcal{V}}

\def\uno{\mathbf{1}}
\def\ie{\emph{i.e. }}

\def\col{\textnormal{col}}
\def\row{\textnormal{row}}
\def\diag{\textnormal{diag}}
\def\tr{\textnormal{tr}}
\def\sp{\textnormal{span}}
\def\st{\textnormal{stack}}
\def\o{\omega}
\def\g{\gamma}

\def\s{\sigma}

\def\xh{\widehat{\bf{x}}}
\def\Xh{\widehat{\bf{X}}}

\def\Et{\widetilde{{\bf{E}}}}
\def\Pt{\widetilde{{P}}}

\def\xb{{\bf{x}}}
\def\yb{{\bf{y}}}

\def\zh{{\bf{z}}}
\def\eb{{\bf{e}}}

\def\fb{{\bf{f}}}
\def\gb{{\bf{g}}}

\def\hb{{\bf{h}}}
\def\hti{{\tilde{\bf{h}}}}
\def\Xb{{\bf{X}}}
\def\Eb{{\bf{E}}}

\def\Xb{{\bf{X}}}
\def\Zb{{\bf{Z}}}

\def\Pinf{P_\infty}

\begin{document}

\title{Consensus analysis of random sub-graphs for distributed filtering with link failures}
\author{Stefano Battilotti\thanks{S. Battilotti is with the Dipartimento Ingegneria Informatica, Automatica e Gestionale (DIAG) "A. Ruberti", Sapienza Università di Roma, Via Ariosto 25, 00185, Rome, Italy  (e-mail: battilotti@diag.uniroma1.it).},
Filippo Cacace\thanks{F. Cacace, is with Universit\`a Campus Bio-Medico di Roma, Rome,  Italy  (e-mail: f.cacace@ieee.org).},
Massimiliano d'Angelo\thanks{M. d'Angelo is with the Dipartimento Ingegneria Informatica, Automatica e Gestionale (DIAG) "A. Ruberti", Sapienza Università di Roma, Via Ariosto 25, 00185, Rome, Italy  (e-mail: mdangelo@diag.uniroma1.it).}}

\maketitle

\begin{abstract}
In this paper we carry out a stability analysis of a distributed consensus algorithm  in presence of link failures. The algorithm combines a new broadcast version of a Push-Sum algorithm, specifically designed for handling link failures,  with a new recursive consensus filter. The analysis is based on the properties of random Laplacian matrices and random sub-graphs and it may also be relevant for other distributed estimation problems. We characterize  the convergence speed, the minimum number of consensus steps needed and the impact of link failures and for both the broadcast Push-Sum and the recursive consensus algorithms. Numerical simulations validate the theoretical analysis.
\end{abstract}

{\bf Keywords}:
Filtering, Network analysis, Stochastic systems, random graphs.

\section{Introduction} \label{sec:intro}

Distributed estimation is based on the usage of multiple sensor nodes to cooperatively
perform large-scale sensing tasks that cannot be accomplished by individual devices.
The availability of low-cost sensors and the diffusion of wireless networks
has contributed in recent years to the development of many applications based on
distributed estimation and filtering, and this area has become
one of the most active topics in filtering theory \cite{ge2020distributed,he2020distributed,battilotti2020asymptotically}.
Distributed filtering algorithms dictate the way in which the information is exchanged and elaborated
by the nodes of the network in order to reach a shared estimate of the target systems state
under the constraint that each node can communicate only with its neighbors, in a broad range
of different settings related to the kind of systems (continuous-time or discrete-time, linear
or non-linear) and communications pattern (bi-directional or uni-directional, broadcast or asynchronous, static or dynamic networks).
In this paper we are concerned with stochastic linear time-invariant systems at discrete-time
on bi-directional links, represented by an undirected graph, when the links are affected
by random and symmetric link failures.

Most of the existing approaches to distributed filtering belong to the consensus-based paradigm
and can be broadly categorized into three groups \cite{he2020distributed}: state estimate fusion \cite{olfati2007distributed,olfati2009kalman,talebi2021stability,ugrinovskii2011distributed,yang2019distributed},  measurement vector fusion \cite{ge2017dynamic,kamgarpour2008convergence,li2018weightedly,olfati2005distributed,wan2018distributed,wu2018distributed}, and information vector fusion
\cite{battistelli2014kullback,kamal2013information,talebi2019distributed,wang2017convergence,wei2018stability}. Hybrid approaches have also been proposed \cite{battistelli2014consensus}.
Diffusion-based filters \cite{cattivelli2010diffusion} descend from diffusion adaptation strategies
\cite{sayed2014diffusion} but they share a structure similar to consensus-based filters.
In the discrete-time setting most of  these approaches are based on multiple consensus steps among neighbors per unit of time, in order to preserve stability of the estimation error and overcome
the problem of non-local observability.
A numerical comparison among some of these methods is reported in \cite{DKF_DT}.
An interesting practical conclusion is that none of the existing methods ensures acceptable estimation accuracy, for arbitrary networks and systems, when too few consensus steps per time unit are used.

Many recent works consider the distributed estimation problem in presence of link failures \cite{ge2020distributed}. Two distinct settings can be considered, since failures can be either  known or unknown to the nodes, \ie the node may either be aware that the message from one of its neighbors got lost or not.
The recent paper \cite{jin2022distributed} proposes distinct filters for the two cases.
We focus on the first case, for the second case see \cite{battilotti2018distributed,jin2022distributed} and the references therein.
The setting that we consider is identical to the one in \cite{Liu2018kalman}. In that proposal  the error covariance does not converge to the centralized Kalman Filter,
the error is not mean square bounded but only stochastically bounded
and the lower bound on the number of consensus steps is not provided.
Analogue limitations affect \cite{jin2022distributed} for the sub-optimal SDKF as
the lower bound on the consensus step cannot be characterized.
An approach based on state estimate fusion with one consensus step is \cite{GaoNiuSheng2022}, where sector-bounded nonlinear systems and directed graphs are considered in the $H_\infty$ framework.
The setting is more general and the results correspondingly weaker: the $H_\infty$ properties of the filter are difficult to compute in a distributed way. Analogue issues arise for the distributed Extended Kalman Filter in \cite{zhu2020hybrid}.
The recent paper \cite{yu2020distributed} extends the approach in \cite{battistelli2014kullback}
to the case of unknown inputs, random link failures and time-varying systems.
The main problems in this case are the  high communication overhead, the performance with few consensus steps and the problem of verifying the  convergence conditions in a distributed environment.

In this paper we propose and analyze an extension of the DKF algorithm of \cite{DKF_DT}
in presence of link failures. DKF makes use of two separate consensus algorithms.
The first one aims at reaching consensus on the overall sensing capability of the network, the second one is a consensus algorithm for the recursive estimation phase.
The use of two separate algorithms allows to reduce the communication burden without losing the convergence to the centralized filter as the number of consensus steps grows.
A feature of DKF in \cite{DKF_DT} is that it is possible to compute at each node the minimum number
of consensus steps that yields mean square boundedness of the estimation error.
We aim at extending these properties in presence of random link failures and
at characterizing the convergence speed, the lower bound on the number of consensus steps
and the impact of link failures.
The second contribution is to propose a modification of the broadcast Push-Sum algorithm to the case of link failures.

The problem is formally defined in Section \ref{sec:problem}.
The distributed filtering  algorithm is presented in Section \ref{sec:dkf_loss} and a separate stability analysis for the two parts of the algorithm is provided in Section \ref{sec:dkf_loss_analysis}. The example in Section \ref{sec:simulations} is used to validate the theoretical results.
Additional results for this analysis are available in the appendices.

\paragraph*{Notation and preliminaries}
$\Nat$, $\Real$ and $\Comp$ denote respectively natural, real and complex numbers.
$|S|$ is the cardinality of set $S$.
For a square matrix $A$, $\tr(A)$ is the trace, $\s(A)$ is the spectrum
and $\rho(A)$ is the spectral radius. $A$ is said to be Schur stable
if $\rho(A)<1$.
$\|A\|$ denotes  the matrix operator norm.
$\E\{\cdot\}$ denotes expectation. $\otimes$ is the Kronecker
product between vectors or matrices, and $M^{[n]}=M\otimes M^{[n-1]}$ with $M^{[0]}=I$
the Kronecker power.
The operators $\row_i()$, $\col_i()$, $\diag_i()$ denote respectively the horizontal, vertical and diagonal compositions of matrices and vectors indexed by $i$. $\st(M)\in\Real^{nm\times1}$ denotes the stack of the matrix $M\in\Real^{n\times m}$. $I_n$ is the identity matrix in $\Real^n$, $1_N\in\Real^N$ is a vector with all entries $1$ and $U_N=1_N1_N^\top$ is the square matrix of size $N$ having $1$ in each entry. A random sequence $\{X_t(\omega)\}$ is asymptotically unbiased if $\lim_{t\to\infty}\E[X_t]=0$, while it is mean square bounded if $\sup_{t\geq 0}\E[\|X_t\|^2]< +\infty$.


\section{Problem statement and assumptions}\label{sec:problem}
On an undirected network connecting $N$ nodes we study the distributed recursive estimation problem of the process
\begin{align}
  \xb_{t+1} =& A\xb_t  + \fb_t ,  \label{eq:system} \\
\yb^{(i)}_t =& C_i \xb_t  + \gb^{(i)}_t , \quad  i=1,\,\dots,\,N, \label{eq:yi}
\end{align}
where $\xb_t  \in \Real^n$, $ \yb^{(i)}_t  \in {\Real^{q_i}}$, $q_i\geq0$, is the output
available at the node $i\in \{1,2 \dots  ,N \}$, $\fb_t $
and $\gb^{(i)}_t$, $i=1,\,\dots,\,N$,  are zero-mean white noises, mutually independent
with covariance respectively $Q$ and $R_i$, $i=1,\ldots, N$.
The matrix $R =  \diag_i(R_i)$ is non-singular.
$\xb_0$ is a random variable with mean ${\widebar{\xb}_0 }:=\E\{ \xb_0\}$ and  covariance
$\Psi_0$.
We denote $\yb_t =\col_i(\yb^{(i)}_t )$ and $C =\col_i(C_i)$ is the aggregate matrix of the output maps.
When $q_i=0$ the node does not have sensing capabilities.
In order to obtain an estimate of $\xb_t$ at each node it is therefore necessary to exchange
information among neighboring nodes.
We consider the problem of designing a distributed state estimator for \eqref{eq:system}--\eqref{eq:yi}
consisting of $N$ local estimators, one for each node, that exchange local information with the neighbors in presence of possible communication failures.

The information exchange between the nodes is modeled by the  undirected graph $\Gc=(\Vc,\Ec)$ where the vertices
$\Vc=\{1,2 \dots  ,N \}$ represent the $N$ nodes and $\Ec \subseteq \Vc \times \Vc$ is the set of edges of the graph.
The presence of an edge $(i,j)$ in ${\Gc}$ implies that nodes $i$ and $j$ can
exchange information between them.
The graph is undirected, that is, the edges $(i,j)$ and $(j,i) \in \Ec$ are considered to be the same.
Two nodes $i$ and $j$, with $i \ne j$, are neighbors to each other if $(i,j)  \in \Ec$.
The set of neighbors of node $i$  is $\Nc^{(i)} := \{  j \in \Vc :(j,i)  \in \Ec  \}$, $\nu^{(i)}=|\Nc^{(i)}|$ its cardinality, $\Ncb^{(i)}=\{i\}\cup\Nc^{(i)}$ is the set set of neighbors including  $i$ itself with $\bar\nu^{(i)}  = \nu^{(i)}+1$.
A path is a sequence of connected edges in a graph. A graph is connected if there is
a path between every pair of vertices.
The adjacency matrix $\Ac\in\Real^{N\times N}$ of  ${\Gc}$ has the $(i,j)$-th entry $1$ if $(i,j)\in\Ec$  and $0$ otherwise.
The degree matrix $\Dc=\diag_{i=1}^N(\nu^{(i)})$ of $\Gc$ is a diagonal matrix
whose $i$-th entry is $\nu^{(i)}$.
The Laplacian of an undirected ${\Gc}$ is the symmetric matrix $\Lcb\in\Real^{N\times N}$ defined by $\Lcb = -\Ac+\Dc$.
When the graph is connected,  $0 = \lambda_1 (\Lcb) < \lambda_2 (\Lcb)\le \cdots \le \lambda_N(\Lcb)$,
where $\lambda_i(\Lcb)$ denotes an eigenvalue of $\Lcb$.
An  eigenvector associated to $ \lambda_1 (\Lcb)$ is $\uno_N $.

\begin{Assumption}\label{ass:connected}
 The graph $\Gc=(\Vc,\Ec)$ is connected.
\end{Assumption}
\begin{Assumption}\label{ass:system}
 The couple $(C,A)$ is observable and the couple $(A,Q^{\frac{1}{2}})$ is controllable.
\end{Assumption}
Assumption \ref{ass:system} ensures that the variance of the estimation error of the centralized
filter is bounded.
The presence of failures is modeled by the Bernoulli stochastic processes $\beta_t^{(ij)}(\o)\in\{0,\,1\}$, where $t$ is the discrete time index. In particular, when $\beta_t^{(ij)}(\o)=1$ node $i$ and $j$ can communicate at time $t$, while the contrary happens when $\beta_t^{(ij)}(\o)=0$.
We assume that these variables have uniform probability $P(\beta_t^{(ij)}(\o)=1)=p_\beta$  and that they are temporally independent and independent from noise $\fb_t$, $\gb^{(i)}_t$.
We set  $\beta_t^{(ii)}(\o) = 1$, since node $i$ has its own measurement, and  we assume that the link failures  are symmetric, \ie if at time $t$ node $i$ does not receive a message from node $j$, the same happens to node $j$ with respect to node $i$.
We also define the process $\beta_{t,h}^{(ij)}(\o)$, with $h\in \Nat$ and the same properties as $\beta_t^{(ij)}(\o)$, to model losses the consensus step iterations. Let $\nu^{(i)}_t(\o)=\sum_{j\in\Nc^{(i)}}\beta_t^{(ij)}(\o)$ be the number of neighbors at time $t$ in presence of  link failures  and  $\bar{\nu}^{(i)}_t(\o)=\nu^{(i)}_t(\o)+1$.
Finally, we introduce the disconnection probability $p_d$, which is the probability, constant
in time, that the network is disconnected at time $t$ due to link failures. $p_d$ depends on $p_\beta$
but also on the the network topology
We summarize these properties in the following assumption.
 \begin{Assumption}\label{ass:loss}
  The disconnection probability is less than $1$, $p_d<1$.
  For all $i,j\in \Vc$, $t\geq 0$, $h\geq 0$, $\beta_t^{(ij)}$, $\beta_{t,h}^{(ij)}$, $\fb_t$, $\gb^{(i)}_t$  are mutually independent. Moreover, for $i\neq j$, the sequences $\beta_t^{(ij)}(\o)$ and $\beta_{t,h}^{(ij)}(\o)$ are identically distributed Bernoulli sequences with mean $p_\beta<1$,  and  $\beta_t^{(ij)}(\o)=\beta_t^{(ji)}(\o)$ and $\beta_{t,h}^{(ij)}(\o)=\beta_{t,h}^{(ji)}(\o)$, while $\beta_t^{(ii)}(\o)\equiv1$.
 \end{Assumption}


\section{Distributed Kalman Filter with link failures}\label{sec:dkf_loss}
The proposed algorithm, shown in Fig. \ref{fig:DKF_lf}, is an extension of the Distributed Kalman filter in \cite{DKF_DT} (DKF) where we integrate the distributed algorithm for the estimation of the gain and the link failures. The initial estimate and covariance of error at node $i$, $\xh_0^{(i)}$ and $\Psi^{(i)}_0$, are chosen as $\xb_0$ and $\Psi_0$ when the values are available.
$\varepsilon$ is an arbitrarily small positive real.
The algorithm has two parameters, the consensus gain $\delta\in\Real$ and the number of
consensus iterations $\gamma\in\Nat$. In detail:

\begin{figure}
\hrule
\textbf{Algorithm DKF with link failures, node $i$}
\hrule
\begin{small}
\begin{enumerate}
 \item [1:] Set \textbf{stop}=false, $\xh^{(i)}_{0\vert 0}= \xh_0^{(i)}$, $P^{(i)}_0=\Psi^{(i)}_0$, $\tilde{C}^{(i)}_0 = C_i^\top R_i^{-1}C_i$, $\tilde{n}^{(i)}_0 =1$, $w^{(i)}_0=0$  except for $w^{(1)}_0=1$ and  $\bar{\nu}^{(i)}= |\Ncb^{(i)} |$.
 \item [2:] At time $t\geq0$, if \textbf{stop} then go to Step 3, otherwise execute:
 \begin{enumerate}
 \item[2.1:] Send $\frac{\tilde{C}^{(i)}_t}{\bar{\nu}^{(i)}}$, $\frac{\tilde{n}^{(i)}_t}{\bar{\nu}^{(i)}}$, $\frac{w^{(i)}_t}{\bar{\nu}^{(i)}}$ to the neighbors.
 \item[2.2:] Compute
\begin{align}
 \hskip-1cm \tilde{C}^{(i)}_{t+1}= \frac{\tilde{C}^{(i)}_t}{\bar{\nu}^{(i)}} &+ \sum_{j\in\Nc^{(i)}}
 \beta_t^{(ij)}\frac{\tilde{C}^{(j)}_t}{\bar{\nu}^{(j)}}+
 (1-  {\beta}_t^{(ij)} ) \frac{\tilde{C}^{(i)}_t}{\bar{\nu}^{(i)}}  \label{eq:zi_lf}  \\
 \hskip-1cm \tilde{n}^{(i)}_{t+1} = \frac{\tilde{n}^{(i)}_t}{\bar{\nu}^{(i)}}&+ \sum_{j\in\Nc^{(i)}}
 \beta_t^{(ij)}\frac{\tilde{n}^{(j)}_t}{\bar{\nu}^{(j)}}+
 (1-  {\beta}_t^{(ij)} ) \frac{\tilde{n}^{(i)}_t}{\bar{\nu}^{(i)}} \\
  \hskip-1cm  w^{(i)}_{t+1}= \frac{w^{(i)}_t}{\bar{\nu}^{(i)}}&+
 \beta_t^{(ij)}\frac{w^{(j)}_t}{\bar{\nu}^{(j)}}+ \sum_{j\in\Nc^{(i)}}
 (1-  {\beta}_t^{(ij)} )  \frac{w^{(i)}_t}{\bar{\nu}^{(i)}}\label{eq:wi_lf} \\
  \hskip-1cm   (G^{(i)}_{t+1},\,N^{(i)}_{t+1}) &= \begin{cases}
            \left(\frac{\tilde{C}^{(i)}_{t+1}}{w^{(i)}_{t+1}},\,\frac{\tilde{n}^{(i)}_{t+1}}{w^{(i)}_{t+1}}\right),\ \textnormal{if } w^{(i)}_{t+1}\neq0, \\
            (C_i^\top R_i^{-1}C_i,\,1)\ \textnormal{otherwise}.
           \end{cases} \label{eq:gi_lf}
\end{align}
 \item[2.3:] If $\|\tilde{C}^{(i)}_{t+1}\|>0$ and $\|\tilde{C}^{(i)}_{t+1}-\tilde{C}^{(i)}_{t}\|<\varepsilon$ then set \textbf{stop}=true.
\end{enumerate}
 \item[3:] Compute
\begin{align}
P^{(i)}_{t+1}=& (AP^{(i)}_{t}A^\top+Q)(I_n+G^{(i)}_{t+1}(AP^{(i)}_{t}A^\top+Q))^{-1} \label{eq:Pi}\\
K^{(i)}_{t+1}=& \begin{cases} N^{(i)}_{t+1} P^{(i)}_{t+1}C_i^\top R_i^{-1}\ \textnormal{if } C_i\neq0, \\ 0 \ \textnormal{otherwise}.
\end{cases}\label{eq:Ki}
\end{align}
\item[4:] Get $\yb^{(i)}_{t+1}$ and compute
\begin{equation*}
    \zh^{(i)}_{t+1,0}= A\xh^{(i)}_{t\vert t} +  K^{(i)}_{t+1} ( \yb^{(i)}_{t+1}- C_i A \xh^{(i)}_{t \vert t}).
\end{equation*}

 \item[5:] For $h=0,\ldots,\g-1$ do
 \begin{enumerate}
 \item[5.1.] Send $\zh^{(i)}_{t+1,h}$ to the neighbors
 \item[5.2.] Compute
\begin{align}
 \zh^{(i)}_{t+1,h+1}=  \zh^{(i)}_{t+1,h} & + \frac{1}{\delta}\sum_{j \in \Nc^{(i)} }\beta^{(ij)}_{t,h} (\zh^{(j)}_{t+1,h}-\zh^{(i)}_{t+1,h}) \label{eq:DKF_00}
 \end{align}
 \end{enumerate}
 \item[6.] Set $\xh^{(i)}_{t+1\vert t+1} =  \zh^{(i)}_{t+1,\gamma}$ and go to Step 2.
\end{enumerate}
\end{small}
\caption{Distributed Kalman Filter (DKF) with link failures.}\label{fig:DKF_lf}
\end{figure}
\begin{itemize}
 \item Step 2 is a modification of the Push-Sum algorithm \cite{DKF_DT,kempe2003gossip} in presence of link failures (see Remark \ref{rem:PS_LF_new}).  We shall prove  that
 \begin{align}
 \frac{\lim_{t\to\infty}\tilde{C}^{(i)}_t}{\lim_{t\to\infty}w^{(i)}_t} =& G= \sum_{i=1}^N C_i^\top R_i^{-1} C_i \\
 \frac{\lim_{t\to\infty}\tilde{n}^{(i)}_t}{\lim_{t\to\infty}w^{(i)}_t}=& N
 \end{align}
and consequently $G^{(i)}_{t} \to \sum_{i=1}^N C_iR_i^{-1}C_i$, $N^{(i)}_{t}\to N$.
\item Eqs. \eqref{eq:Pi}--\eqref{eq:Ki} compute the filter gain. Considering
the asymptotic values of $G^{(i)}$, $N^{(i)}$ given above it is easy to check that
$K^{(i)}_t\to N \Pinf C_i^\top R_i^{-1}$, where $\Pinf$ is the solution of the
Riccati equation for the centralized Kalman filter (\ie the Kalman filter
with all $\bf{y}_t$ available),
 \begin{align}
P_\infty  & = (I_n-K_\infty C) (A P_\infty A^\top+Q)(I_n-K_\infty C)^\top
\nonumber\\
&\quad + P_\infty C^\top R^{-1} C P_\infty, \label{eq:ric_CKF}\\
K_\infty &= P_\infty C^\top R^{-1}. \label{eq:Kinf}
\end{align}
\item Steps 4 and 5 compute the local state estimate.
Steps 4 implements a local prediction and correction step with the locally available measurements.
Step 5 is a dynamic averaging in $\gamma$ steps with gain $\delta$ on the estimates.
\end{itemize}
\begin{Remark}\label{rem:PS_LF_new}
The original Push-Sum algorithm \cite{kempe2003gossip} is a gossip-based
protocol to compute aggregate functions over static networks.
Broadcast versions, more suited to networks of dynamical agents, have been proposed in \cite{DKF_DT} and \cite{shames2012distributed}. The Broadcast Push-Sum algorithm computes
$\sum_{i=1}^N x^{(i)}$ as $y^{(i)}_0=x^{(i)}$, $w_0^{(i)}=0$, $w_0^{(1)}=1$,
    \begin{align}\label{eq:ps}
    y^{(i)}_{t+1}= \frac{y^{(i)}_t}{\bar{\nu}^{(i)}}+ \sum_{j\in\Nc^{(i)}}
 \frac{y^{(j)}_t}{\bar{\nu}^{(j)}}, \quad
 w^{(i)}_{t+1}= \frac{w^{(i)}_t}{\bar{\nu}^{(i)}}+ \sum_{j\in\Nc^{(i)}}
 \frac{w^{(j)}_t}{\bar{\nu}^{(j)}}.
    \end{align}
$w^{(i)}_t$ is a positive scalar weight.
In \cite{DKF_DT} it is proved that $y^{(i)}_t/w^{(i)}_t$ converges to $\sum_{i=1}^N x^{(i)}$.
The algorithm used here is a new version to deal with link failures.
In both the gossip and broadcast versions each node receives from its neighbors the consensus data divided by the number of its neighbors.
This cannot be done in a synchronized way because, due to failures,
the nodes do not know the instantaneous number of neighbors when transmitting.
The new algorithm does not use the variable number of neighbors $\bar{\nu}^{(i)}_t(\o)$
but the nominal (constant) value $\bar{\nu}^{(i)}$, replacing missing packets,
$\beta^{(ij)}_t=0$, with local values, namely
$\tilde{C}^{(i)}_t/\bar{\nu}^{(i)}$, $\tilde{n}^{(i)}_t/\bar{\nu}^{(i)}$, $w^{(i)}_t/\bar{\nu}^{(i)}$.
\end{Remark}
\begin{Remark}
 It is worth noticing that, differently from other popular consensus algorithms,
 $\tilde{C}^{(i)}_t-\tilde{C}^{(j)}_t$  does not converge to zero, and the asymptotic values of $\tilde{C}^{(i)}_t$, $\tilde{n}^{(i)}_t$ and $w^{(i)}_t$
 depend on the agent.
 However, $\tilde{C}^{(i)}_t/\bar{\nu}^{(i)}_t-\tilde{C}^{(j)}_t/\bar{\nu}^{(j)}_t\to0$ for all  $(i,j)$.
 The estimates of the sums of interest, $N=\sum_{i=1}^N \tilde{n}^{(i)}$ and
 $G=\sum_{i=1}^N C_i^\top R_i^{-1}C_i$,  are obtained as $\tilde{n}^{(i)}_t/w^{(i)}_t$ and $\tilde{C}^{(i)}_t/w^{(i)}_t$.
\end{Remark}
\begin{Remark}
    When it is challenging to orchestrate the initial conditions $w^{(i)}_0$ to be zero
    for all the nodes apart from one it is possible to resort to the \emph{max-consensus}
    algorithm of \cite{shames2012distributed} in order to effectively decide a leader in the
    graph. The analysis of the resulting algorithm in presence of link failures can be
    developed in analogy with the one reported here.
\end{Remark}
\begin{Remark}
We assume that in absence of link failures the network graph $\Gc$ is static.
In this condition, the first part of the algorithm, Step 2,
can be stopped when the asymptotic values of $G^{(i)}$, $N^{(i)}$ and $K^{(i)}$ are reached.
Moreover, under standard hypothesis, the iterative Riccati equation at Step 3 reaches its steady-state
value in a limited number of additional iterations.
Thus, at steady state, the filter reduces at Steps 4,\,5 of Fig. \ref{fig:DKF_lf}.
\end{Remark}

\begin{Remark}
Failures are modeled by separate and independent variables.
 At each time $t>0$ link failures that occur at Step 2.2,
 \eqref{eq:zi_lf}, \eqref{eq:wi_lf}, are modeled by $\beta_t^{(ij)}(\o)$, and those in the averaging
 Step 5.2 are modeled by $\beta_{t,h}^{(ij)}(\o)$.
\end{Remark}

\begin{Remark}
 The choice of the parameters $\delta$ and $\gamma$ will be considered in the next section.
 This choice has no influence on Steps 2, 3.
\end{Remark}

\section{Analysis of the DKF with link failures}\label{sec:dkf_loss_analysis}

The estimation algorithm in Fig. \ref{fig:DKF_lf} can be broadly
divided in two parts. Steps 2,\,3 aim at estimating the optimal gains
$K^{(i)}$ by computing $G=C^\top R^{-1}C=\sum_{i=1}^N C_i^\top R_i^{-1}C_i$,
whereas the second one uses $K^{(i)}$ for the online estimation algorithm.
For this reason we divide our analysis in two parts.
We first show that $K^{(i)}\to N \Pinf C_i^\top R_i^{-1}$. We can then analyze the behavior of the estimation algorithm when $K^{(i)}$ is replaced by its asymptotic value.

The communication pattern at time $t$ (resp. $(t,h)$ for the dynamic averaging)
is described by the random Laplacian matrix
\begin{equation}\label{eq:stoch_Lap}
 (\Lc_{t}(\omega))_{i,j}=\begin{cases}
                    -\beta_t^{(ij)}(\o),\textnormal{ if } i\neq j\\
                    \nu^{(i)}_t(\o)=\sum_{\ell\in\Nc^{(i)}} \beta_t^{(i\ell)}(\o),\textnormal{ if } i=j.
                   \end{cases}
\end{equation}
Assumption \ref{ass:loss} guarantees that the random Laplacian still enjoys
the properties of the initial Laplacian matrix $\Lcb$.
\begin{Lemma}\label{lem:Lomega}
 If Assumption \ref{ass:loss} holds then: (i) $\E[\Lc_t(\omega)]= p_\beta\Lcb$;
(ii) $\Lc_t(\omega)$ is symmetric and positive semi-definite; (iii) $0\in\s(\Lc_t(\o))$
and the associated left and right normalized eigenvectors are $v=\frac{1}{\sqrt{N}}\uno_N$ and $v^\top$;
(iv) for all $\omega$, $\rho(\Lc_t(\omega))\leq\rho(\Lcb)$.
\end{Lemma}

\textit{Proof}. The first three properties are a consequence of the fact that $\Lc_t(\omega)$
is still the Laplacian matrix of an undirected graph. To prove $\rho(\Lc_t(\omega))\leq\rho(\Lcb)$
we recall the following well known property (see \cite{chung1997spectral}, Section 1.2),
$ x^\top\Lc x = \sum_{(i,j)\in\Ec} (x_i-x_j)^2$,
that implies, $\forall x\in\Real^N$, $0\leq x^\top\Lc_t(\omega)x \leq x^\top \Lcb x$,
that is, $\|\Lc_t(\omega)\|\leq\|\Lcb\|$. Both matrices are symmetric and positive
semi-definite, and then $\max(\sigma(\Lc_t(\omega)))\leq\max(\sigma(\Lcb))$. \hfill$\Box$

\subsection{Convergence analysis of the local gains}\label{ssec:dkf_loss_Ki}

The goal of this section is to prove that in presence of link failures the broadcast
Push-Sum algorithm at Step 2 in Fig. \ref{fig:DKF_lf} computes at each node $N$ and
$G=\sum_{i=1}^NC_i^\top R_i^{-1}C_i$.
This provides a distributed computation of $\Pinf$ and $K^{(i)}$ that are essential
for the filters equation at Steps 4,\,5 of DKF.

The convergence of the broadcast Push-Sum algorithm has already been studied \cite{shames2012distributed}. We report for completeness the convergence result
for the case without link failures.

\begin{Theorem}\label{th:pushsum_static}
In absence of link failures (\ie $\beta_t^{(ij)}=\beta_{t,h}^{(ij)} = 1$\, $\forall t,h\geq0$, $i\neq j$), if Assumption \ref{ass:connected} holds then the sequences $\tilde{C}^{(i)}_t$, $\tilde{n}^{(i)}_t$, $w^{(i)}_t$ generated by \eqref{eq:zi_lf}--\eqref{eq:wi_lf} are such that
\begin{align}
 \lim_{t\to\infty} \tilde{C}^{(i)}_t\to & \frac{\bar{\nu}^{(i)}}{\sum_{j=1}^N\bar{\nu}^{(j)}}\sum_{j=1}^N C_j^\top R_j^{-1}C_j \label{eq:limz}\\
 \lim_{t\to\infty} \tilde{n}^{(i)}_t\to & \frac{\bar{\nu}^{(i)}}{\sum_{j=1}^N\bar{\nu}^{(j)}} N\label{eq:limn}\\
 \lim_{t\to\infty} w^{(i)}_t\to & \frac{\bar{\nu}^{(i)}}{\sum_{j=1}^N\bar{\nu}^{(j)}} \label{eq:limw}
\end{align}
\end{Theorem}

The result is proved in Appendix \ref{app:pushsum_static}.
The extension to the case of link failures, which is one of the main results of this paper,  requires Assumption \ref{ass:loss}.

\begin{Theorem}\label{th:pushsum_lf}
If Assumption \ref{ass:connected} and Assumption \ref{ass:loss} are satisfied
then the limits \eqref{eq:limz}--\eqref{eq:limw} hold in probability.
\end{Theorem}

The proof is reported in Appendix \ref{app:pushsum_lf}.

\subsection{Asymptotic analysis of the estimation errors}\label{ssec:dkf_loss_analysis}

In this section we analyze the performance of the DKF with link failures,
corresponding to Steps 4,\,5 in Fig. \ref{fig:DKF_lf}. In particular, we prove that the estimate provided by DKF with link failures is asymptotically unbiased and mean square bounded for a sufficiently large $\delta$ and $\gamma$. This  provides also some guidelines to tune the parameters $\delta$ and $\gamma$.
We assume that the values of $G^{(i)}$ and $N^{(i)}$ have already
reached their respective asymptotic value, $G=C^\top R^{-1}C$ and $N$, and $K^{(i)}=N\Pinf C_i^\top R_i^{-1}$ where $\Pinf$ solves \eqref{eq:ric_CKF}.

Let $\Xh_t=\col_{i=1}^N(\xh^{(i)}_{t|t})\in\Real^{nN}$ be the estimates
at the nodes, and $\Xb_t= \uno_N\otimes \xb_t\in\Real^{nN}$ the vector corresponding to the
system state that evolves as
\begin{equation}
 \Xb_{t+1}= \uno_N\otimes(A\xb_t+\fb_t)= (I_N\otimes A)\Xb_t+ \uno_N\otimes\fb_t. \label{eq:Xt1}
\end{equation}
The overall estimation error is thus $\Eb_t= \Xb_t-\Xh_t=\col_{i=1}^N(\eb^{(i)}_t)$,
with $\eb^{(i)}_t=\xb_t-\xh^{(i)}_{t|t}$.

Let $\Lcb=\Lcb^\top$ denote the communication graph in absence of failures.
The spectrum of $\Lcb$ is $\sigma(\Lcb)=\{\lambda_1=0,\lambda_2,\ldots,\lambda_N\}$,
where $\lambda_i\in\Real$ for $i=1,\ldots,N$. Let
\begin{align}
 M_t^{(\gamma)}(\omega)=& \prod_{h=0}^{\g-1}\left( I_N - \frac{1}{\delta}\Lc_{t,h}(\omega)\right)\in\Real^{N\times N}.\label{eq:Mgamma}
\end{align}
The averaging Step 5.2 can be represented through $M_t^{(\gamma)}(\o)$ in \eqref{eq:Mgamma}
and $\Zb_{t+1,h}=\col_{i=1}^N(\zh^{(i)}_{t+1,h})$, $i=1,\ldots,N$, as
\begin{equation}
\Zb_{t+1,h+1} = \left(M_t^{(\g)}(\o)\otimes I_n\right)\Zb_{t+1,h}.
 \end{equation}
Let $A_i=(I-K^{(i)}C_i)A$.
From \eqref{eq:ric_CKF}, \eqref{eq:Kinf} it follows that, for $A_C=A-K_\infty CA$.
\begin{align}
\frac{1}{N}\sum_{i=1}^N A_i =&
 \frac{1}{N}\sum_{i=1}^N (I_n-K^{(i)}C_i)A= (I_n- K_\infty C)A= A_C.
\end{align}
We therefore have
\begin{align}
 \zh^{(i)}_{t+1,0}=& A\xh^{(i)}_{t|t}+ K^{(i)}\left( C_iA\left(\xb_t-\xh^{(i)}_{t|t}\right)
 + C_i\fb_t+\gb^{(i)}_{t+1}\right)\nonumber\\
 =& A\xb_t- A_i\eb^{(i)}_t+  K^{(i)} \left( C_i\fb_t+\gb^{(i)}_{t+1}\right) \\
 \Xh_{t+1}=& \left(M_t^{(\g)}(\o)\otimes I_n\right)(\uno_N\otimes A\xb_t)
 \nonumber\\
 & - \left(M_t^{(\g)}(\o)\otimes I_n\right)\diag_i(A_i) \Eb_t \nonumber \\
 & + \left(M_t^{(\g)}(\o)\otimes I_n\right)\col_i\left(\ K^{(i)}(C_i\fb_t+\gb^{(i)}_{t+1}) \right) \nonumber\\
 =& (I_N\otimes A)\Xb_t - \left(M_t^{(\g)}(\o)\otimes I_n\right)\diag_i(A_i) \Eb_t \nonumber \\
 & + \left(M_t^{(\g)}(\o)\otimes I_n\right)\col_i\left(\ K^{(i)}(C_i\fb_t+\gb^{(i)}_{t+1}) \right),\label{eq:Xh1}
\end{align}
where we have used the property
\begin{align*}
 \left(M_t^{(\g)}(\o)\otimes I_n\right)(\uno_N\otimes A\xb_t)= (\uno_N\otimes A\xb_t)= (I_N\otimes A)\Xb_t
\end{align*}
that descends from $M_t^{(\g)}(\o)\uno_N=\uno_N$ (see Lemma \ref{lem:Mtgamma}, point 3, in Appendix \ref{app:aux}).
By subtracting \eqref{eq:Xh1} from \eqref{eq:Xt1}
\begin{align}
 \Eb_{t+1} &= A_\Gc^{(\g)}(\o)\Eb_t+ \left(M_t^{(\g)}(\o)\otimes I_n\right)\col_{i=1}^N(\hb^{(i)}_t),\label{eq:Et1}
\end{align}
with
\begin{align}
A_\Gc^{(\g)}(\o) &= \left(M_t^{(\g)}(\o)\otimes I_n\right)\diag_i(A_i) \label{eq:Agomega}\\
\hb^{(i)}_t &=  (I_n-K^{(i)}C_i)\fb_t-K^{(i)}\gb^{(i)}_{t+1}. \label{eq:ht}
\end{align}
The stability analysis of $\Eb_t$ in \eqref{eq:Et1} is not trivial, because $A_\Gc^{(\g)}(\o)$
is a random matrix. Our analysis makes use of Lemma \ref{lem:random_stability} in Appendix \ref{app:random_matrices}
that provides sufficient conditions for mean square boundedness of linear systems with random matrices.
We also use the following simple bound.
\begin{Lemma}\label{lem:Erho2}
 Let $M(\o)\in\Real^{n\times n}$ be a symmetric random matrix such that $\forall\o:\ \rho(M(\o))\leq1$
 and $\P(\{\o:\rho(M)=1\})<1$. Then $\rho(\E[M(\o)])<1$.
\end{Lemma}
\textit{Proof}. Since $M(\o)$ is symmetric, $\rho(M(\o))=\|M(\o)\|$ and $\rho(\E[M(\o)])=\|\E[M(\o)\|$.
Clearly, $1>\E[\rho(M(\o))]$, and by Jensen's inequality
$ 1>\E[\rho(M(\o))] = \E[\|M(\o)\|]\geq \quad  \|\E[M(\o)]\|= \rho(\E[M(\o)])$.
\hfill$\Box$

The matrix $M_t^{(\g)}(\o)$ in \eqref{eq:Et1} is in general not symmetric but it enjoys several
interesting properties.
\begin{Lemma}\label{lem:eigMbargamma}
The expected value of $M_t^{(\gamma)}(\o)$ is
\begin{align}
\E\left[M_t^{(\gamma)}(\o)\right] =&
\bar{M}^{(\gamma)}= \left( I_N - \frac{p_\beta}{\delta}\Lcb\right)^\gamma.\label{eq:Mbargamma}
\end{align}
 The eigenvalues of $\bar{M}^{(\gamma)}$ are $\{ (1-\frac{p_\beta}{\delta}\lambda_i)^\gamma\}$,
 with $\lambda_i\in\sigma(\Lcb)$. In particular, $1\in\sigma(\bar{M}^{(\gamma)})$.
\end{Lemma}
The proof is reported in \ref{app:proof_eigMbargamma} together with other auxiliary results
and properties.
The main result is that the estimate provided by DKF with link failures is asymptotically unbiased and mean square bounded for a sufficiently large $\delta$ and $\gamma$ (the proof is reported in Appendix \ref{app:dkf_unbiased}).
\begin{Theorem}\label{th:dkf_unbiased}
 When Assumption \ref{ass:connected}, \ref{ass:system} and \ref{ass:loss} hold
 and $\delta>\rho(\Lcb)$ there exists $\g_0\in\Nat$
 such that $\forall\g>\g_0$ the estimate of the DKF filter (Steps 4 and 5 in Fig. \ref{fig:DKF_lf}) is asymptotically  unbiased and mean square bounded.
\end{Theorem}

A particular cases arises when the output matrices are identical, \ie $C_i=C_j$, $i,j=1,\ldots,N$,
that implies $A_i=A_C$ for all $i =1,\ldots,N$.
In this case, any number of consensus steps yields asymptotic unbiased estimates.

\begin{Corollary}\label{cor:special}
 In the hypotheses of Theorem \ref{th:dkf_unbiased}, if in addition $C_i=C_j$ for $i,j=1,\ldots,N$
 then for any $\g\geq1$ the estimate of the DKF filter (Steps 4 and 5 in Fig. \ref{fig:DKF_lf}) is asymptotically  unbiased and mean square bounded.
 Moreover, in this case the equation \eqref{eq:Et1}
 of the estimation error in the transformed coordinates $\Et_t=\Tc^{-1} \Eb_t=\Tc^\top\Eb_t$, where $\Tc$ is defined in \eqref{eq:change_of_coord} is
 \begin{equation}
  \Et_{t+1}= \begin{pmatrix}
     A_C & 0 \\ 0 & S_t^{(\g)}(\o) \otimes A_C
    \end{pmatrix}\Et_t + \tilde{\hb}_t, \label{eq:Etilde1_special}
 \end{equation}
where $S_t^{(\g)}(\o)$ is defined in \eqref{eq:Stgamma} and $\tilde{\hb}_t$ in \eqref{eq:hbt}.
\end{Corollary}
The proof is reported in Appendix \ref{app:cor_special} and
the properties of the matrix $S_t^{(\g)}(\o)$ are summarized in Lemma \ref{lem:Mtgamma}.
The following lemma provides the asymptotic value of the covariance of the transformed noise $\hti_t$ as $\g\to\infty$.

\begin{Lemma}\label{lem:tilde_h}
Let $\Psi_{\hti}=\E[\hti_t\hti_t^\top]\in\Real^{nN\times nN}$.
 If Assumption \ref{ass:connected}, \ref{ass:system} and \ref{ass:loss} hold
 and $\delta>\rho(\Lcb)$ then $\lim_{\g\to\infty} \Psi_{\hti}$ is equal to
 \begin{equation*}
   \begin{pmatrix} N(I_n-K_\infty C)Q(I_n-K_\infty C)^\top+ NP_\infty C^\top R^{-1}CP_\infty & 0 \\
  0 & 0
  \end{pmatrix}.
 \end{equation*}
\end{Lemma}
The proof of this lemma is reported in Appendix \ref{app:lem_tilde_h}.
We are now in the position of extending to the case of link failures another key property of DKF.

\begin{Theorem}\label{th:cov_err_i}
 If Assumption \ref{ass:connected}, \ref{ass:system} and \ref{ass:loss} hold  and $\delta>\rho(\Lcb)$ then
 the covariance of the estimation error at each node tends for $t\to\infty$ and $\g\to\infty$ to the covariance $P_\infty$  of the centralized Kalman filter.
\end{Theorem}
The proof is reported in Appendix \ref{app:th_cov_err_i}.

All the results have been obtained under the condition $\delta>\rho(\Lcb)$, guaranteeing
that $\E[S_t^{(\g)}(\o)^{[2]}]$ and $\bar{S}^{\g}$ are Schur.
Since the eigenvalues of $\bar{S}$ are $1-\frac{p_\beta}{\delta}\lambda_i$
with $\lambda_i\in\sigma(\Lcb)>0$, there is trade-off between stability and performance.
A large $\delta$ is safer but the convergence
to $0$ of the expected value of the estimation error is slow and the variance can be large.
Thus, a value only slightly larger than $\rho(\Lcb)$ is in practice the best choice.
Clearly, $\delta>\rho(\Lcb)$ cannot be verified by using only local information, but
one can use $\delta>\max_i\{\nu^{(i)}\}>\rho(\Lcb)$ (see \cite{DKF_DT}), where
 $\max_i\{\nu^{(i)}\}$ can be computed through a distributed algorithm.

\subsection{Bounds for the number of consensus steps}\label{ssec:gamma_bounds}

A lower bound of $\gamma$ for the mean dynamics $\E[\Eb_t]$ can be
obtained by guaranteeing the error system
\begin{align}
\E[\Eb_{t+1}] =& \E[A_\Gc^{(\g)}] \E[\Eb_t]= \left(\bar{M}^{(\g)} \otimes I_n\right)\diag_i(A_i)\E[\Eb_t]   \label{auxsys}
\end{align}
 is asymptotically stable. To this aim, we will reason on the dual global error system $\eb_{t}= \E\left[A_\Gc^{(\g)}\right]^\top \eb_{t}$. If
\begin{equation}
\Tc^\top  \eb_{t} = ( T^\top \otimes I_n)    \eb_{t}  =:   {\widetilde \eb}_{t }   =  \begin{pmatrix}  {\widetilde \eb}_{t}^{(1)}   \cr {\widetilde \eb}_{t }^{(2)} \end{pmatrix},
\end{equation}
 where $\Tc=(v\ W)\otimes I_n$ as in \eqref{eq:change_of_coord},
  we obtain
\begin{align}
\begin{pmatrix} {\widetilde \eb}_{t+1}^{(1)} \cr   {\widetilde \eb}_{t+1}^{(2)} \end{pmatrix}     =& \E[H_t(\omega)]^\top \begin{pmatrix} {\widetilde \eb}_{t}^{(1)} \cr   {\widetilde \eb}_{t}^{(2)} \end{pmatrix}= \bar{H}^\top \begin{pmatrix} {\widetilde \eb}_{t}^{(1)} \cr   {\widetilde \eb}_{t}^{(2)} \end{pmatrix},
\end{align}
with $H_t(\omega)$ defined in \eqref{eq:Htom}, $\bar{H}=\E[H_t]$,
\begin{align}
\bar{H}^\top =&\begin{pmatrix}  A_C^\top & \E[{H}_{21}^{(\g)}]^\top \\ H_{12}^\top
 & \E[{H}_{22}^{(\g)}]^\top  \end{pmatrix}   =
 \Tc^\top {\widebar A}_D^\top \Tc \begin{pmatrix} I_n & 0 \\ 0 & \bar{S}^\g\otimes I_n \end{pmatrix}  \label{io}
\end{align}
where $\bar{S}=\E[S_{t,h}(\omega)]$ is defined in \eqref{eq:Sbar},
${H_{21}^{(\g)}}(\omega)$, ${H_{22}^{(\g)}}(\omega)$ defined in \eqref{eq:Htom},
${\widebar A}_D=\diag_{i=1}^N(A_i)$. The Riccati equation \eqref{eq:ric_CKF} guarantees
that it is always possible to pick $\lambda \in (0,1)$  such that
\begin{equation}\label{eq:lyap_pinf}
A_C P_\infty A_C^\top  \le     \lambda P_\infty.
\end{equation}
In the proof of the next theorem we use the weighted norms
\begin{equation}
\Vert N  \Vert_{M} := \sup_{z \in {\mathbb R}^m} \sqrt{   \frac{z^\top N^\top M N z}{z^\top M z} } ,
\end{equation}
$N \in\Real^{m \times m}$ , $M\in{\mathcal P}_+ (m)$.
\eqref{eq:lyap_pinf} implies $\|A_C^\top\|_{P_\infty}\leq\sqrt{\lambda}$.
\begin{Theorem}\label{prop:lower_mean}
If $\delta>\rho(\Lcb)$ the DKF with link failures is asymptotically unbiased for all $\g$ such that
$\rho(A_R(\gamma))<1$,
\begin{align}
  A_R(\gamma) :=& \begin{pmatrix} \sqrt{\lambda} & \theta_{p_\beta}^\gamma c_B
  \\ c_B & \theta_{p_\beta}^{\gamma} c_B \end{pmatrix} \in\Real^{2\times 2}\label{eq:bb} \\
 c_B := & (1+ N\| P_\infty\| \|G\|_{P_\infty})\|A\|_{P_\infty} , \label{eq:bb1}
\end{align}
where $\lambda\in(0,1)$ satisfies \eqref{eq:lyap_pinf} and $\theta_{p_\beta} := 1-\frac{p_\beta}{\delta}(1-\cos(\pi/N))<1$.
\end{Theorem}
\textit{Proof}.
 Since $\|\Tc\|=1$ and $\|{\widebar A}_D^\top\|=\|\diag_{i}(A_i)\|=\max_i\|A_i\|$,
\begin{align}
 \Vert \Tc^\top {\widebar A}_D^\top \Tc \Vert_{I_N \otimes P_\infty}
& = \max_i\|A_i\|_{I_N \otimes  P_\infty}\leq c_B. \label{aux2}
\end{align}
Moreover, notice that, with $\theta_{p_\beta}$ defined in \eqref{eq:normSbar},
\begin{align}
\Vert \bar{S}  \otimes I_n   \Vert_{I_{N-1} \otimes  P_\infty} =&\|\bar{S}  \otimes I_n \| \leq \theta_{p_\beta}, \label{eq:theta}\\
  \Vert (\bar{S}  \otimes I_n )^\g \Vert_{I_{N-1} \otimes P_\infty} =&  \Vert  \bar{S} \Vert^\g  \leq \theta_{p_\beta}^\g.
\end{align}
By using  $\|A_C^\top\|_{P_\infty}\leq\sqrt{\lambda}$,  (\ref{io}) and (\ref{aux2}) we have
\begin{align}
 \Vert {\widetilde \eb}_{t+1}^{(1)}\Vert_{P_\infty} \le &
 \sqrt{\lambda} \Vert {\widetilde \eb}_{t}^{(1)} \Vert_{P_\infty} + \theta_{p_\beta}^\gamma c_B  \Vert    {\widetilde \eb}_{t}^{(2)} \Vert_{I_{N-1} \otimes  P_\infty} \\
\Vert {\widetilde \eb}_{t+1}^{(2)} \Vert_{I_{N-1} \otimes  P_\infty}  \leq &
 \,c_B   \Vert {\widetilde \eb}_{t}^{(1)} \Vert_{P_\infty} +  \theta_{p_\beta}^\gamma c_B  \Vert    {\widetilde \eb}_{t}^{(2)} \Vert_{I_{N-1} \otimes  P_\infty}
\end{align}
Finally, we conclude
\begin{align}
 \begin{pmatrix}  \Vert {\widetilde \eb}_{t+1}^{(1)}  \Vert_{P_\infty}  \\ \Vert  {\widetilde \eb}_{t}^{(2)}  \Vert_{I_{N-1} \otimes  P_\infty}   \end{pmatrix}  \le &  A_R(\g)
 \begin{pmatrix}  \Vert {\widetilde \eb}_{t}^{(1)}  \Vert_{   P_\infty}  \cr   \Vert  {\widetilde \eb}_{t}^{(2)}  \Vert_{I_{N-1} \otimes  P_\infty}   \end{pmatrix} \label{eq:proofTh}.
\end{align}
Thus, for the stability of  (\ref{auxsys}), it is sufficient that  the spectral radius of $ A_R (\g)  $ is less than $1$.
\hfill $\Box$
\begin{Corollary}
    The condition \eqref{eq:bb} is always satisfied when
    \begin{equation}\label{eq:bb_new}
        \left( 1- \frac{p_\beta(1-\cos{(\pi/N)})}{\delta}\right)^\gamma
        < \left(\frac{1-\sqrt{\lambda}}{c_B}\right)^2
    \end{equation}
\end{Corollary}
The proof of the bound of the lower bound on  $\gamma$ expressed by \eqref{eq:bb_new}
is obtained by computing an upper bound of $\rho(A_R(\g))$ in \eqref{eq:bb} less than $1$.

\begin{figure*}[th]
\begin{align}
\E\left[H_t(\omega)^{[2]}\right] =
 \left(\begin{array}{@{}c|c c c@{}}
    A_C^{[2]} & A_C\otimes H_{12} & H_{12}\otimes A_C & {H_{12}}^{[2]} \\
    \hline
    A_C \otimes \E[{H}_{21}^{(\g)}] & A_C \otimes \E[{H}_{22}^{(\g)}] &
    H_{12}\otimes \E[{H}_{21}^{(\g)}] & H_{12}\otimes \E[{H}_{22}^{(\g)}] \\
    \E[{H}_{21}^{(\g)}]\otimes A_C & \E[{H}_{21}^{(\g)}]\otimes H_{12} &
     \E[{H}_{22}^{(\g)}]\otimes A_C & \E[{H}_{22}^{(\g)}]\otimes H_{12} \\
    \E[{H_{21}^{(\g)}}^{[2]}] & \E[H_{21}^{(\g)} \otimes H_{22}^{(\g)}] &
    \E[H_{22}^{(\g)} \otimes H_{21}^{(\g)}] & \E[{H_{22}^{(\g)}}^{[2]}]
\end{array}\right) \nonumber
\end{align}
\caption{Structure of the matrix $\E\left[H_t(\omega)^{[2]}\right]$.}\label{fig:EH2tom}
\end{figure*}
\begin{Remark}
    The parameter $\theta_{p_\beta}$ defined in Theorem \ref{prop:lower_mean} depends
     on the probability of successful communications, that can be estimated from each
     node from the statistics of the (known) link failures. Since for all $p_\beta<1$ we have
      $\theta_{p_\beta}<1$, there is always $\gamma$ such that $\rho(A_R(\gamma))<1$.
     The matrices $P$, $G$, $N$ and $A$ are known to each node, thus the lower bound on $\gamma$ can be
     locally computed.
\end{Remark}

\begin{Theorem}\label{th:mean_square_boundedness}
If $\delta>\rho(\Lcb)$ the DKF with link failures is mean square bounded for all $\g$ such that
$\rho(A_S(\gamma))<1$,
\begin{align}
  A_S(\gamma) :=& \begin{pmatrix} \lambda &  c_B \sqrt{\lambda} & c_B\sqrt{\lambda} & c_B^2 \\
  c_B\theta_{p_\beta}^{\gamma}\sqrt{\lambda} &  c_B\theta_{p_\beta}^{\gamma}\sqrt{\lambda}
  & c_B^2 \theta_{p_\beta}^{\gamma} & c_B^2 \theta_{p_\beta}^{\gamma} \\
  c_B\theta_{p_\beta}^{\gamma} \sqrt{\lambda} & c_B^2 \theta_{p_\beta}^{\gamma}
  & c_B\theta_{p_\beta}^{\gamma} \sqrt{\lambda}& c_B^2 \theta_{p_\beta}^{\gamma} \\
  c_B^2 \theta_{p_d}^{\gamma} & c_B^2 \theta_{p_d}^{\gamma} & c_B^2 \theta_{p_d}^{\gamma} & c_B^2 \theta_{p_d}^{\gamma}
  \end{pmatrix}\label{eq:bb2}
\end{align}
where $\lambda\in(0,1)$ satisfies \eqref{eq:lyap_pinf}, $c_B$ is defined in \eqref{eq:bb1},  $\theta_{p_\beta}<1$ is defined in \eqref{eq:normSbar} and $\theta_{p_d}<1$ is defined in Theorem \ref{prop:lower_mean}.
\end{Theorem}
\textit{Proof}.
After the change of coordinates $\Et_t=\Tc^\top\Eb_t$ that leads to
\eqref{eq:Etilde1} we derive the conditions for mean square bounded
through $\rho(\E[H_t(\o)^{[2]}])<1$ (see Lemma \ref{lem:random_stability}, iii).
The matrix $\E[H_t(\o)^{[2]}]$ is shown in Fig. \ref{fig:EH2tom}.
By using the same approach as in Theorem \ref{prop:lower_mean} we derive an upper bound
for the spectral radius with the spectral radius of the matrix in $\Real^{4\times4}$
obtained by using upper bounds for the norms of the blocks of $\E[H_t(\o)^{[2]}]$.
The upper bounds for the norms of the blocks of the first three rows are the same as in
Theorem \ref{prop:lower_mean}. For the fourth row we need an upper bound for
$\|\E[{S_t^{(\g)}(\o)}^{[2]}]\|$, which is provided by Lemma \ref{lem:Mtgamma}, 2), (iii).
\hfill $\Box$
\begin{Remark}
    The bound \eqref{eq:bb2} depends not only on the link probability $p_\beta$
    but also on the disconnection probability $p_d$. This last parameter depends
    on $p_\beta$ and the network topology. Thus the agents must have some information
    on the network configuration in order to estimate $p_d$.
\end{Remark}

\section{Simulation results}\label{sec:simulations}
\begin{figure}[b]
\centering
 \includegraphics[scale=0.2]{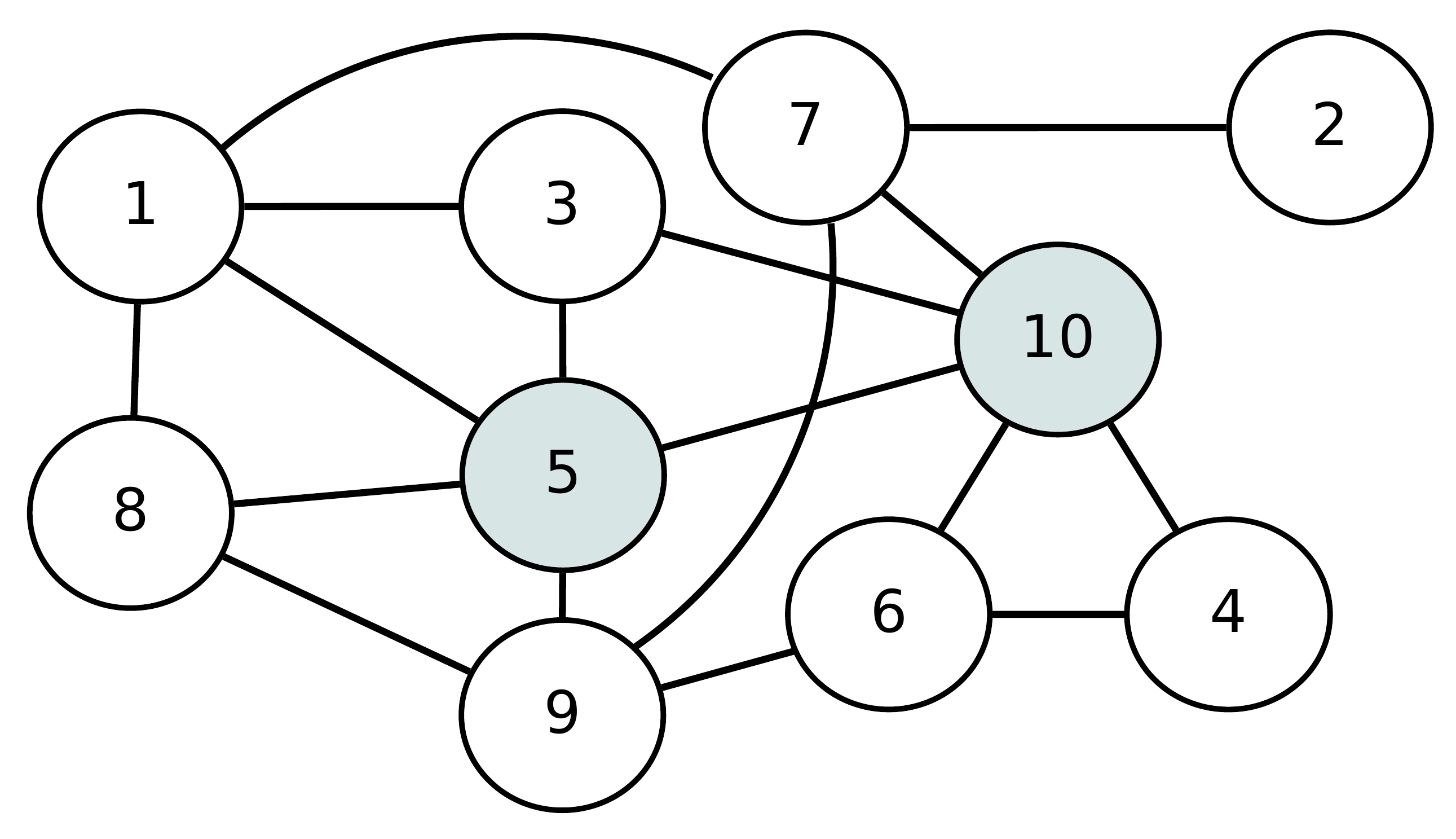}
 \caption{Network topology of the example.}\label{fig:graph}
\end{figure}


\begin{figure*}[ht]
\centering
 \includegraphics[scale=0.85]{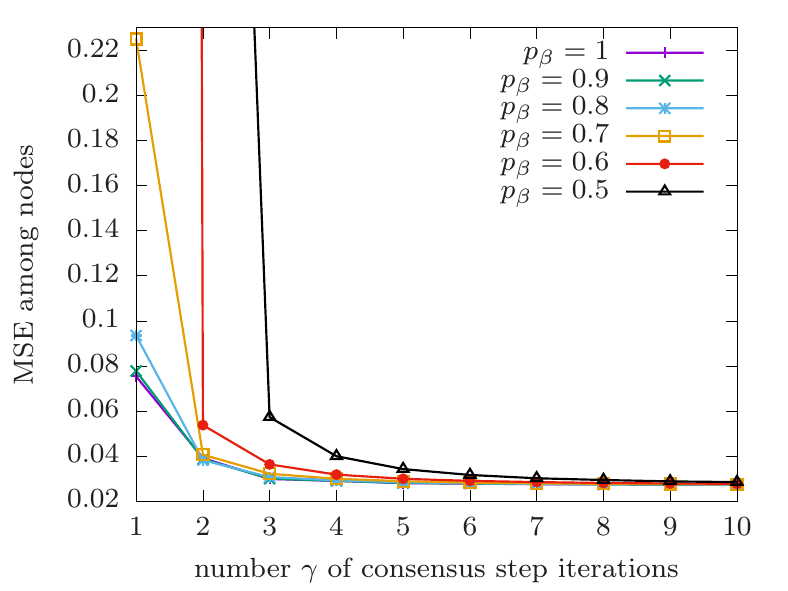}
 \includegraphics[scale=0.85]{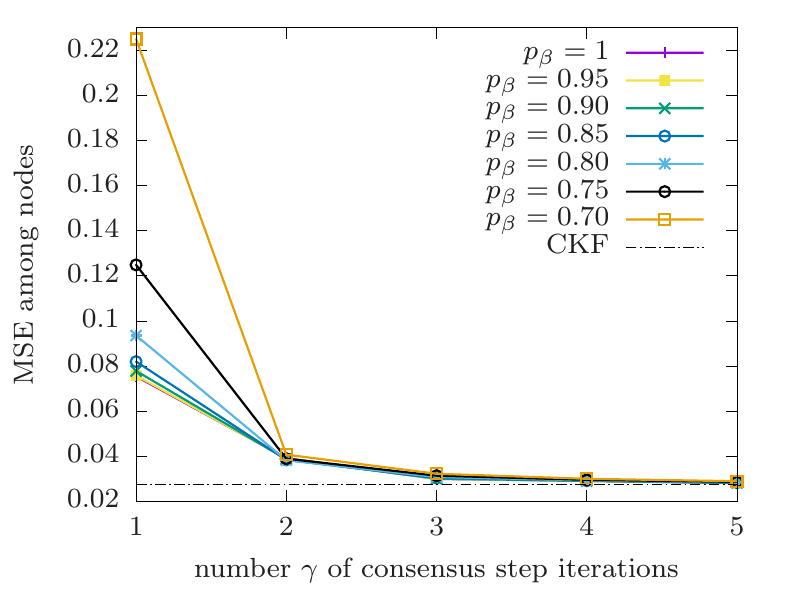}
  \caption{MSE among nodes with respect to $\g$ for different values of probability of receiving information, namely $p_\beta$. The plot on the right is an enlargement of the left plot with a finer discretization of the probability levels.}\label{fig:mse}
\end{figure*}

\begin{figure}[ht]
\centering
\includegraphics[scale=1]{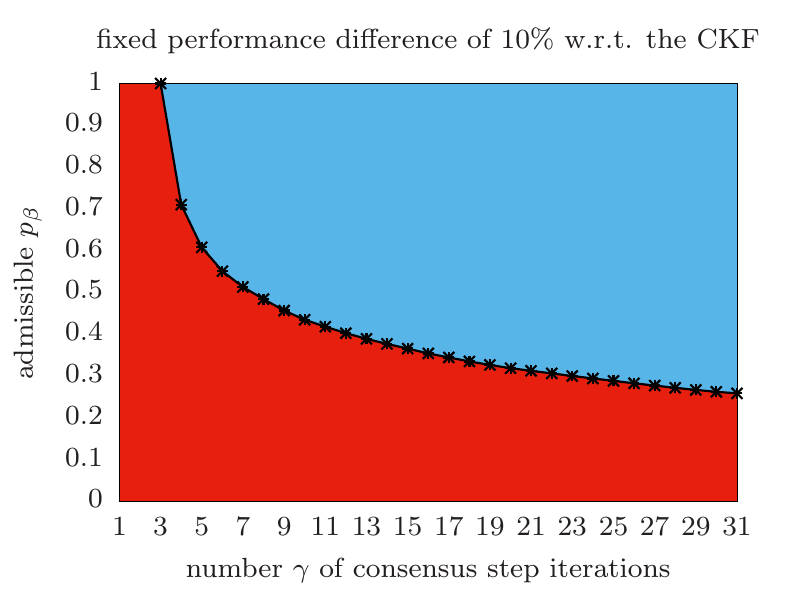} 
 \caption{Minimum admissible value of probability $p_\beta$ of receiving information with respect to the number $\g$ of consensus step iterations in order to keep the performance difference (in terms of MSE) with respect to the centralized Kalman filter (CKF) without link failures around $10\%$. The red area denotes the region where performance difference degrades over $10\%$, while the pale blue area denotes the region where performance difference are below $10\%$.}\label{fig:maxp}
\end{figure}

We consider the problem of tracking a planar system with a discretization step $\tau$, where the sensors may estimate only one component of the position, namely
\begin{align}
 A=&\begin{pmatrix}
    I_2 & \tau I_2 \\ 0 & I_2
   \end{pmatrix},\quad C_1=\begin{pmatrix}1 & 0 & 0 & 0\end{pmatrix}
   \quad R_i=\s^2_g \nonumber \\
   Q=&\s^2\begin{pmatrix}
      \frac{1}{3}\tau^3I_2 & \frac{1}{2}\tau^2I_2 \\ \frac{1}{2}\tau^2I_2  & I_2
     \end{pmatrix},\quad C_2=\begin{pmatrix}0 & 1 & 0 & 0\end{pmatrix}.\nonumber
\end{align}
We consider the topology of Fig. \ref{fig:graph} consisting of a connected graph with $N=10$ nodes and $17$ arcs. Nodes $5$ and $10$ have measurement matrices $C_1$
and $C_2$ respectively, the remaining nodes have communication capabilities only.
With $\tau=0.25$, $\s=0.05$, $\s_g=0.1$ we performed
$N_s=300$ simulations of $\bar{k}=450$ points for several values of $\g$.
The MSE is computed as the  norm of the estimation error averaged over all the nodes and times $k\in[\frac{1}{5}\bar{k},\dots,\bar{k}]$  (to avoid transient effects). Fig.~\ref{fig:mse} show the MSE averaged over all the nodes with respect to the number $\g$ of consensus step iterations for different values of probability of link failure, namely $1 - p_\beta$. The plot on the right is an enlargement of the left plot with a finer discretization of the probability levels and the line corresponding to CKF is the performance of the centralized Kalman filter without link failures. We can see that the proposed filter is still stable with $\g = 1$ with acceptable performance for a probability of link failure $1-p_\beta = 0.3$, whilst the  MSE of the filter is unbounded for probability of link failure $1-p_\beta = 0.4$ with $\gamma=1$ and probability of link failure $1-p_\beta = 0.5$ with $\gamma\in\{1,2\}$. As expected, when $\g$ increases the performance of the filter improves and tends to the centralized optimal one with no link failure (Fig.~\ref{fig:mse}, right). Fig.~\ref{fig:mse} (right) shows that for $p_\beta \in [0.7, 1]$ and $\g\geq 2$ the performance are essentially the same.
Fig.~\ref{fig:maxp} shows for the considered example the minimum value $p_\beta$ of receiving information with respect to the number $\g$ of consensus step iterations in order to keep MSE within a difference of  $10\%$ with respect to the centralized Kalman filter (CKF, no link failures). The red area denotes the region where the performance difference degrades over $10\%$ while the pale blue area denotes the region where the performance difference is below $10\%$. We observe that in this example it is not possible to keep the performance difference with respect to the CKF without link failures within $10\%$ if $\g\leq 2$.

\section{Conclusions}\label{sec:conclusions}

This article studies the stability of a distributed filter with multiple consensus steps
in presence of random link failures when the system is linear time-invariant,
failures are symmetric and not correlated in time.
The extension to directed simmetrizable graphs can be carried out  as in \cite{DKF_DT}.
Future developments may aim at extending the framework to time-varying or non-linear
systems, random delays \cite{battilotti2019stochastic}, non-symmetric and correlated fault processes or attacks \cite{guan2017distributed}.

\appendix
\section{Proofs}\label{app:proofs}

\subsection{Proof of Lemma \ref{lem:eigMbargamma}}\label{app:proof_eigMbargamma}

For the independence of the stochastic Laplacians,
\begin{align}
 \E[M_t^{(\gamma)}(\omega)]&= \prod_{h=0}^{\g-1}\left( I_N-\frac{1}{\delta} \E[\Lc_{t,h}(\omega)]\right)
= \bar{M}^{(\gamma)}.
\end{align}
Let $U$ be the non-singular matrix such that $\Lcb= U\Lambda U^{-1}$, where
$\Lambda$ is diagonal and real. Then,
\begin{align}
 \bar{M}^{(\gamma)} = \left( I_N - \frac{p_\beta}{\delta}\Lcb\right)^\gamma &=
 U \left( I_N- \frac{p_\beta}{\delta}\Lambda\right)^\gamma U^{-1}.
\end{align}
Clearly, the eigenvalues of $I_N- \frac{p_\beta}{\delta}\Lambda$ are $\{ 1-\frac{p_\beta}{\delta}\lambda_i\}$, with left and right eigenvectors $\{e_j\}$, the versors of the canonical base.
It follows from $(I_N- \frac{p_\beta}{\delta}\Lambda)e_j= (1-\frac{p_\beta}{\delta}\lambda_j)e_j$ that
$(I_N- \frac{p_\beta}{\delta}\Lambda)^\gamma e_j= (1-\frac{p_\beta}{\delta}\lambda_j)^\gamma e_j$,
and, due to the similarity relationship, $(1-\frac{p_\beta}{\delta}\lambda_j)^\gamma\in\sigma(\bar{M}^{(\gamma)})$.
Since in particular $\lambda_1=0$, $1\in\sigma(\bar{M}^{(\gamma)})$.
\hfill$\Box$

\subsection{Auxiliary results}\label{app:aux}

\begin{Lemma}\label{lem:pushsum}
Let $\Ac$ and $\Dc$ be, respectively, the adjacency and degree matrix of an undirected connected graph.
Then, matrix $\Theta=(I+\Ac)(I+\Dc)^{-1}$ has the following properties:
\begin{itemize}
 \item[a)] $\Theta$ is non-negative, irreducible and column stochastic.
 \item[b)] $\rho(\Theta)=1=\lambda_1\in\sigma(\Theta)$, and $\lambda_1=1$ is simple.
 \item[c)] $\Theta$ is primitive, that is, $\lambda_1=1$ is the only eigenvalue on the unit circle, \ie $|\lambda_i|<1$ for $i\neq1$.
 \item[d)] The right and left eigenvectors $u_1$, $v_1^\top$  of $\lambda_1=1$ are the only
 non-negative eigenvectors, where
 \begin{align}
  u_1& = \col_{i=1}^N(\bar{\nu}^{(i)})= \col_{i=1}^N(\nu^{(i)}+1) \label{eq:u1}\\
  v_1^\top &= \uno_N^\top.\label{eq:v1}
 \end{align}
\end{itemize}
\end{Lemma}
\textit{Proof.}
Point a). By construction, the entries of $\Theta$ are either $0$ or positive. The irreducibility
of $\Theta$ follows from the fact that the matrix $\Ac$ is the adjacency matrix of a connected and
undirected graph. From the property
\begin{align*}
  \Theta =& (I_N+\Dc +\Ac -\Dc)(I_N+\Dc)^{-1}= I_N-\Lc (I_N+\Dc)^{-1}
\end{align*}
it is easy to check that the entries of the $i$-th column of $\Theta$
are either $0$ or $1/({\nu}^{(i)}+1)$, \ie the non-zero entries are  identical and there
are ${\nu}^{(i)}+1$ of them, so each column sums up to $1$.

Point b). The property $\rho(\Theta)\in\sigma(\Theta)$ follows from the Perron-Frobenius theorem for non-negative irreducible matrices (see for example \cite{meyer2000matrix}, ch. 8). An application of the Gerschgorin Circles to the columns of $\Theta$ yields $\rho(\Theta)\leq1$. We show that $\lambda_1=1\in\sigma(\Theta)$ and therefore $\rho(\Theta)=1$.
In fact
\begin{align*}
 \Theta\cdot u_1&= \Theta\cdot \col_{i=1}^N(\bar{\nu}^{(i)})
 = (I_N+\Ac)\cdot\uno_N=\col_{i=1}^N(\bar{\nu}^{(i)}).
\end{align*}

Point c). $\Theta$ is primitive since it is a non-negative irriducible matrix with non-zero diagonal entries, which is a sufficient condition for primitivity (\cite{meyer2000matrix}, ch. 8).

Point d). We have already shown that $u_1=\col_{i=1}^N(\bar{\nu}^{(i)})$ is the right eigenvector
of $\lambda_1=1$. Moreover, since the columns to $\Theta$ sum up to $1$,
$v_1^\top\cdot \Theta= \uno_N^\top\cdot \Theta=\uno_N^\top= v_1^\top$.
Finally, $u_1$ and $v_1^\top$ are the only nonnegative eigenvectors as stated by the Perron-Frobenius theorem.
\hfill $\Box$

\begin{Lemma}\label{lem:M1}
If $\delta>\rho(\Lcb)$ then, $\forall\omega$, $\sigma (M_t^{(\gamma=1)}(\omega))\subset[-1,\,1]$,
and $1\in\sigma(M_t^{(\gamma=1)}(\omega))$ with right and left normalized eigenvectors $v=\frac{1}{\sqrt{N}}\uno_N$ and $v^\top$.
\end{Lemma}

\textit{Proof}. Lemma \ref{lem:Lomega} implies that
$\lambda_N(\frac{1}{\delta}\Lc_t(\omega))\leq\lambda_N(\frac{1}{\delta}\Lcb)<1$.
Since $\frac{1}{\delta}\Lc_t(\omega)$ is symmetric and positive semi-definite, by using the same
derivations as in Lemma \ref{lem:eigMbargamma} we obtain that the eigenvalues
of $M_t^{(\gamma=1)}(\omega)= I_N- \frac{1}{\delta}\Lc_t(\omega)$ are $\{ 1- \frac{1}{\delta}\lambda_i(\Lc_t(\omega))\}$,
with $\left| 1- \frac{1}{\delta}\lambda_i(\Lc_t(\omega))\right|\leq1$.
In particular, since $\lambda_1(\Lc_t(\omega))=0$, $1\in\sigma(M_t^{(\gamma=1)}(\omega))$.
\hfill$\Box$

\begin{Lemma}\label{lem:Somega}
If $\Lc_t(\omega)$ is the Laplacian matrix of a connected subgraph of $\Gc$,  $\delta>\rho(\Lcb)$ and
  $W=\row_{i=1}^{N-1}(w_i)\in\Real^{N\times(N-1)}$ is any (deterministic) matrix
  whose columns are an orthornormal base of the subspace $\Sc^\bot_v$ orthogonal
  to  $\Sc_v=\sp(\{v\})$, $v=\frac{1}{\sqrt{N}}\uno_N$, then the matrix
  \begin{equation}\label{eq:Somega}
   S_t(\omega)= I_{N-1}-\frac{1}{\delta}W^\top \Lc_t(\omega)W,
  \end{equation}
  is symmetric, semi-definite positive and Schur.
  Moreover,
  \begin{equation}
  \s(S_t(\omega))=\s(M_t^{(\gamma=1)}(\omega))\setminus\{1\}.
  \end{equation}
\end{Lemma}

\textit{Proof}.
$S_t(\omega)$ is symmetric because $\Lc_t(\omega)$ is symmetric.
$v$ and $\Sc_v$ are independent of $\omega$, hence
it exists a deterministic matrix $W\in\Real^{N\times(N-1)}$ such that
$v^\top W=W^\top v=0$, $W^\top W=I_{N-1}$. The matrices $vv^\top,\,WW^\top\in\Real^{N\times N}$
are the projectors on $\Sc_v$ and $\Sc^\bot_v$ respectively, and $I_N= vv^\top+WW^\top$.
Further, for all $\omega$  and $x\in\Real^N$, $\Lc_t(\omega)x\in \Sc_v^\top$, since $v^\top \Lc_t(\omega)x=0$.
Consequently, $WW^\top \Lc_t(\omega)W= \Lc_t(\omega) W$,
because the columns of $\Lc_t(\omega) W$ are in $\Sc_v^\bot$ and $WW^\top$ is the projector on $\Sc^\bot_v$.
Thus,
\begin{align}
 M_t^{(\gamma=1)}(\omega)&= I_N - \frac{1}{\delta}\Lc_t(\omega) = vv^\top + W S_t(\omega)W^\top. \label{eq:decomposition}
\end{align}
The two matrices of these decomposition are orthogonal.
When $\Lc_t(\omega)$ is connected, the eigenvalue $1\in\sigma(M_t^{(\gamma=1)}(\omega))$
has multiplicity one and it is the only non-null eigenvalue of $vv^\top$ since $vv^\top v= v$.
The remaining $N-1$ eigenvalues of $M_t^{(\gamma=1)}(\omega)$, that according to Lemma \ref{lem:M1} belong to
the interval $(-1,\,1)$, are the eigenvalues of the second term, that is, the eigenvalues of
$S_t(\omega)$ and $0$. \hfill$\Box$

\begin{Lemma}\label{lem:Sbar}
 If $\Lcb$ is the Laplacian matrix of a connected graph of $\Gc$,  $\delta>\rho(\Lcb)$,
  $W$ as in Lemma \ref{lem:Somega}, then the matrix
  \begin{equation}\label{eq:Sbar}
   \bar{S}= I_{N-1}-\frac{p_\beta}{\delta}W^\top \Lcb W,
  \end{equation}
  is symmetric, semi-definite positive and Schur. Moreover,
  \begin{align}
  \|\bar{S}\| &= \rho(\bar{S}) = 1-\frac{p_\beta}{\delta}\lambda_2(\Lcb) \nonumber \\
 & \leq \theta_{p_\beta} := 1-\frac{p_\beta}{\delta}(1-\cos(\pi/N)) \label{eq:normSbar} \\
   \bar{M}^{(\g)}&= vv^\top + W\bar{S}^{\g} W^\top \label{eq:Sbar2}\\
    \bar{S}^{\g}&=  W^\top \bar{M}^{(\g)} W. \label{eq:Sbar3}
  \end{align}
\end{Lemma}
\textit{Proof.} The properties of $\bar{S}$ defined in \eqref{eq:Sbar} are easily proved by the same steps as in Lemma \ref{lem:Somega}, since $\delta/p_\beta\geq\delta>\rho(\Lcb)$.
\eqref{eq:normSbar} follows from \eqref{eq:Sbar} by using the lower bound
$\lambda_2(\Lcb)\geq 2(1-\cos(\pi/N))$ (see \cite{horn2012matrix}).
From \eqref{eq:decomposition} it is easy to see that the identity \eqref{eq:Sbar2} holds for $\g=1$. Proceeding inductively,
\begin{align}
\bar{M}^{(\g)} &= \bar{M}^{(\g-1)} \bar{M}^{(1)} = vv^\top + W\bar{S}^{\g} W^\top.
\end{align}
To obtain \eqref{eq:Sbar3} it is sufficient to pre- and post-multiply
the terms in \eqref{eq:Sbar2} by $W^\top$ and $W$ respectively and recall that $W^\top W=I_{N-1}$.
\hfill$\Box$

\begin{Lemma}\label{lem:Mtgamma}
  If Assumption \ref{ass:loss} holds and
\begin{equation}\label{eq:Stgamma}
 S_t^{(\g)}(\o) := \prod_{h=0}^{\g-1} S_{t,h}(\o).
\end{equation}
  then, for all $t$:
 \begin{enumerate}
  \item
$\E[S_t^{(\g)}(\omega)]=\bar{S}^{\g}$.
  \item If in addition $\delta>\rho(\Lcb)$ then:
  \begin{enumerate}
      \item[(i)] $\E[\|S_t^{(\g)}(\omega){S_t^{(\g)}}(\o)^\top\|]<1$.
      \item[(ii)] $\E[S_t^{(\g)}(\o){S_t^{(\g)}}(\o)^\top]$ is Schur.
      \item[(iii)] $\|\E[{S_t^{(\g)}(\o)}^{[2]}]\| \leq  \theta_{p_d}^\gamma$,
       \begin{align}
         \theta_{p_d}:= (1-p_d)\theta_{p_\beta}^2+ p_d & < 1 \label{eq:theta_p_d}
       \end{align}
      \item[(iv)] $\lim_{\g\to\infty}\|\E[{S_t^{(\g)}(\o)}^{[2]}]\|=0$.
  \end{enumerate}
  \item $1\in\s(M_t^{(\g)}(\o))$ for all $\omega$, and the associated left and right normalized
eigenvectors  are $v=\frac{1}{\sqrt{N}}\uno_N$ and $v^\top$.
 \item if $\delta>\rho(\Lcb)$ and $p_d<1$ then
 \begin{align}
  \lim_{\g\to\infty} \bar{M}^{(\g)}= \lim_{\g\to\infty} \E\left[M_t^{(\g)}(\o)\right]= vv^\top &= \frac{1}{N}U_N \nonumber\\
  \lim_{\g\to\infty} \E\left[{M_t^{(\g)}(\o)}^\top M_t^{(\g)}(\o)\right]= vv^\top&=\frac{1}{N}U_N \nonumber
 \end{align}
\end{enumerate}
\end{Lemma}
\textit{Proof}.
Point 1.
By using the decomposition \eqref{eq:decomposition}, we have
\begin{align}
 M^{(\g)}_t(\omega)&= vv^\top+ W \left(\prod_{h=0}^{\g-1}S_{t,h}(\omega) \right)W^\top. \label{eq:MS}
\end{align}
By taking expectations with $\E[M_t^{(\gamma)}(\omega)]=\bar{M}^{(\gamma)}$ and by comparing
with \eqref{eq:Sbar2} we conclude that $\E[S_t^{(\g)}(\omega)]=\bar{S}^{\g}$.

Point 2.
$S_{t,h}(\o)$ is symmetric and, if $\delta>\rho(\Lcb)$, $\rho(S_{t,h}(\o))\leq1$ (Lemma \ref{lem:Somega}), thus $\|S_{t,h}(\o)\|\leq1$, $\forall\o$.
\begin{align}
\|S_t^{(\g)}(\omega){S_t}^{(\gamma)}&(\omega)^\top\|= \| S_{t,0}(\o)\cdot\ldots\cdot {S_{t,\g-1}}(\o)^2 \cdot\ldots\cdot S_{t,0}(\o)\| \nonumber \\
\leq &\| S_{t,0}(\o)\|^2\cdot\ldots\cdot \|S_{t,\g-1}(\o)\|^2 \leq 1.
\label{eq:normSS}
\end{align}
The equality holds only when $\o$ is such that $\Lc_{t,h}(\o)$
is not connected, $h=0,\ldots,\g-1$, that is, with probability $p_d<1$, thus $\E[\|S_t^{(\g)}(\omega){S_t}^{(\gamma)}(\omega)^\top\|]<1$.
By using \eqref{eq:normSS} and Lemma \ref{lem:Erho2} it follows that $\E[S_t^{(\g)}(\omega){S_t}^{(\gamma)}(\omega)^\top]$ is Schur.
(iii):
\begin{align*}
    \| \E[{{S_t}^{(\gamma)}}^{[2]}]\| =&  \| \E[ (\Pi_{h=0}^{\g-1} S_{t,h})^{[2]} ]\|
    = \| \E[ \Pi_{h=0}^{\g-1} S_{t,h}^{[2]} ] \| \\
   =& \|  \Pi_{h=0}^{\g-1}\E[ S_{t,h}^{[2]} ] \|
   \leq  \prod_{h=0}^{\gamma-1} \| \E[{S_{t,h}}^{[2]}]\| \\
    \leq & \prod_{h=0}^{\gamma-1}  \E[\|S_{t,h}^{[2]}\|]
    =  \E [ \rho(S_{t,h}^{[2]}) ]^\gamma = \E [ \rho(S_{t,h})^{2} ]^\gamma
\end{align*}
At each $(t,h)$ the graph is either connected, with probability $1-p_d$, and
$\rho(S_{t,h})\leq \theta_{p_\beta}$, or disconnected with probability $p_d$
and $\rho(S_{t,h})=1$. It follows $\E [ \rho(S_{t,h})^{2} ] \leq \theta_{p_d}<1$
and iv).

Point 3.
\begin{align}
 M_t^{(\gamma)}(\omega)v&= M_t^{(\gamma-1)}(\o)\left(I_N\cdot v- \frac{1}{\delta}\Lc_{t,h}(\o)v\right)\nonumber\\
 &=M_t^{(\g-1)}(\o)v=\ldots=v.
\end{align}
Point 4. If $\delta>\rho(\Lcb)$,
$\sigma(\E[M_t^{(\gamma)}(\omega)])=\sigma(\bar{M}^{(\gamma)})=\{ (1-\frac{p_\beta}{\delta}\lambda_i)^\gamma\}\subset[-1,\,1]$, where we have used Lemma \ref{lem:eigMbargamma}.
Recalling that $1\in\sigma(\bar{M}^{(\gamma)})$, we obtain that
$\lim_{\g\to\infty}\sigma(\E[M_t^{(\gamma)}(\omega)])=\{0,1\}$.
Thus, $\E[M_t^{(\gamma)}(\omega)]$ tends to the projector on the auto-space of $1$, that is, $vv^\top$.
Moreover, from \eqref{eq:MS} and the property
\begin{equation*}
 \E[A(\omega)B(\omega)B^\top(\omega)A^\top(\omega)]=
\E[A(\omega)\E[B(\omega)B^\top(\omega)]A^\top(\omega)]
\end{equation*}
that holds for any couple of independent stochastic matrices $A(\omega)$, $B(\omega)$ with compatible dimensions, we obtain
\begin{align*}
&\lim_{\g\to\infty}  \E\left[{M_t^{(\g)}}^\top M_t^{(\g)}\right]= vv^\top+ \lim_{\g\to\infty} W\cdot \nonumber\\
&  \cdot\E\Big[ S_{t,0} \E\big[S_{t,1} \ldots
\E\big[ S_{t,\g-1}^2\big] \ldots S_{t,1}\big] S_{t,0}\Big] W^\top = vv^\top.
\end{align*}
In fact, for any $h=0,\ldots,\g-1$, the absolute value of all the eigenvalues of $S_{t,h}(\omega)^2$
are less than $1$ with probability $>0$. Thus, $\E[S_{t,h}(\omega)^2]$ is Schur (Lemma \ref{lem:Erho2})
and the second term of the decomposition goes to $0$ as $\gamma\to\infty$.
\hfill$\Box$

\subsection{Proof of Theorem \ref{th:pushsum_static} }\label{app:pushsum_static}

When $\beta^{(ij)}_t\equiv1$, the recursive equations in \eqref{eq:limz}, \eqref{eq:limn} and \eqref{eq:limw} are of the kind
\begin{align}\label{eq:generic_ps}
 z_i(t+1)= \frac{z_i(t)}{\bar{\nu}^{(i)}}+ \sum_{j_\in\Nc^{(i)}} \frac{z_j(t)}{\bar{\nu}^{(j)}}.
\end{align}
To prove Theorem \ref{th:pushsum_static} we show that for any initial assignment $z_i(0)=x_i$ at each node the sequence \eqref{eq:generic_ps} is such that
\begin{equation}\label{eq:lim_static}
 \lim_{t\to\infty}z_i(t)= \frac{\bar{\nu}^{(i)}}{\sum_{j=1}^N \bar{\nu}^{(j)}}\sum_{j=1}^N x_j,
\end{equation}
since the thesis follows by taking into account the initial assignments for $\tilde{C}^{(i)}$, $\tilde{n}^{(i)}$ and $w_i$.
Let us consider the vector case $x_i\in\Real^n$, the matrix case being a trivial extension. Denoting $Z= \col_{i=1}^N(z_i)\in\Real^{nN}$ \eqref{eq:generic_ps} becomes
\begin{align}
Z(t+1)&= \left((I_N+\Ac)(I_N+\Dc)^{-1}\otimes I_n\right)Z(t)\nonumber \\
& = (\Theta\otimes I_n) Z(t) \label{eq:zt}
\end{align}
with $Z(0)=\col_{j=1}^N x_j$.
Since $(\Theta\otimes I_n)^t= \Theta^t\otimes I_n$,
$\sigma(\Theta^t\otimes I_n)=\sigma(\Theta^t)=\{ \lambda_i^t:\  \lambda_i\in\sigma(\Theta)\}$.
As proved in Lemma \ref{lem:pushsum} the matrix $\Theta$ is non-negative, irreducible and primitive with
spectral radius $\rho(\Theta)=1$, and $\lambda_1=1$ is the only eigenvalue on the unit circle, and then
\begin{align}
&\lim_{t\to\infty}(\Theta\otimes I_n)^t Z(0) =\left(\frac{u_1v_1^\top}{v_1^\top u_1}\otimes I_n\right)\col_{j=1}^N(x_j) \nonumber \\
&= \col_{i=1}^N \left( \frac{\bar{\nu}^{(i)}}{\sum_{j=1}^N\bar{\nu}^{(j)}}\sum_{j=1}^N x_j  \right),
 \label{eq:detZt}
\end{align}
because $u_1=\col_{i=1}^N(\bar{\nu}^{(i)})$ and $v_1^\top=\uno_N$,
thus $(v_1^\top\otimes I_n)\col_{j=1}^N(x_j)= \sum_{j=1}^N x_j$ . \hfill $\Box$

\subsection{Proof of Theorem \ref{th:pushsum_lf} }\label{app:pushsum_lf}

Our aim is to show that with the property \eqref{eq:lim_static} still holds in probability.
In compact form the recursive equations in \eqref{eq:limz}, \eqref{eq:limn} and \eqref{eq:limw} are of the kind
\begin{align}
Z(t+1)&= \left(((I_N+\Ac(t,\o)+ \Dc- \Dc(t,\o)) \right. \nonumber\\
&\cdot \left.(I_N+\Dc)^{-1})\otimes I_n\right)Z(t)\nonumber\\
&= \left( ( I_N -\Lc(t,\o)(I+\Dc)^{-1}) \otimes I_n\right) Z(t) \nonumber\\
&= (\Theta(t,\o)\otimes I_n) Z(t)  \label{eq:zt_lf},
\end{align}
In \eqref{eq:zt_lf}, the matrices $\Ac(t,\omega)$ and $\Dc(t,\omega)=\diag_i(\nu^{(i)}_t)$ are, respectively the (random) adjacency matrix and degree matrix of the random subgraph $\Gc(t,\omega)$ of $\Gc$ that models the symmetric link failures. The random matrix $\Theta(t,\o)\in\Real^{N\times N}$ still enjoys all the properties of $\Theta$ in Lemma \ref{lem:pushsum}, except being irreducible and primitive. The algebraic multiplicity of the eigenvalue $\lambda_1=1$ can be more than one when the graph is disconnected due to link failures, but $\Theta(t,\o)\geq0$, $\Theta(t,\o)$ is column stochastic, $\rho(\Theta(t,\o))=1$, $\|\Theta(t,\o)\|_1=1$.
This last property ensures that the solution of \eqref{eq:zt_lf} is simply stable. Moreover, when $p_\beta>0$, the matrix
\begin{equation}
    \E_\Theta= \E\left[\Theta(t,\o)\right]= I_N- p_\beta\Lcb (I_N+D)^{-1}
\end{equation}
is non-negative, irreducible, primitive and column stochastic with right and left eigenvectors
$u_1$, $v_1^\top$ defined in \eqref{eq:u1}--\eqref{eq:v1}.
Consequently,
\begin{equation}
    \lim_{t\to\infty} \E_\Theta^t= \frac{u_1 v_1^\top}{v_1^\top u_1}= \mathcal{P}_1,
\end{equation}
where we denote $\mathcal{P}_1$ the projector on the autospace of $\lambda_1=1$.
The expected value of $Z(t)$ has limit
\begin{align}
    \lim_{t\to\infty} Z(t)=& \lim_{t\to\infty} \left( \E_\Theta^t \otimes I_n\right)Z(0)
    = \left( \mathcal{P}_1 \otimes I_n\right) Z(0),
\end{align}
identical to the deterministic case \eqref{eq:detZt}.
The proof is concluded by showing that the covariance of $Z(t)$ is asymptotically vanishing.
Suppose, for conciseness, that $n=1$. Then,
\begin{align}
 Z(t)=& \prod_{\tau=0}^{t-1} \Theta(\tau,\omega) Z(0) \\
 \Psi_Z(t) =& \E\left[ \left(\prod_{\tau=0}^{t-1} \Theta(\tau,\o)- \E_\theta^t\right)Z(0)\right. \nonumber\\
 &\left. \cdot Z^\top(0)
 \left(\prod_{\tau=0}^{t-1} \Theta^\top(t-1-\tau,\o)- {\E_\theta^\top}^t\right)\right].
\end{align}
By using the property $\st(AXA^\top)=A^{[2]}\st(X)$ and the temporal independence of $\Theta(\tau,\o)$, denoting $\Psi_0=\st(Z(0)Z^\top(0)$ we arrive at
\begin{align}
    \st(\Psi_Z(t))=& \left(\left(\E\left[ \Theta^{[2]}(t,\o)\right]\right)^t- \left(\E_\Theta^{[2]}\right)^t\right)\Psi_0.
\end{align}
Since the spectrum of the Kronecker square is composed by the product of eigenvalues,
$\s\left(\E\left[ \Theta^{[2]}(t,\o)\right]\right)\subset [0,1]$ and when $p_\beta>0$
the only eigenvalue on the unit circle is $\lambda_1^2=1$, thus
\begin{equation}
    \lim_{t\to\infty}\left(\E\left[ \Theta^{[2]}(t,\o)\right]\right)^t
    =  \lim_{t\to\infty} \left(\E_\Theta^{[2]}\right)^t= \mathcal{P}_1^{[2]},
\end{equation}
and, consequently, $\st(\Psi_Z(t))\to0$. When $n>1$ it is sufficient to replace
$\Theta(\tau,\o)$ with $\Theta(\tau,\o)\otimes I_n$ and $\E_\Theta$ with $\E_\Theta\otimes I_n$
and the proof is identical.

\subsection{Proof of Theorem \ref{th:dkf_unbiased} }\label{app:dkf_unbiased}
Let $T =(v\ W)$, $T_v=v\otimes I_n$, $T_w=W\otimes I_n$ and
\begin{align}\label{eq:change_of_coord}
 \Tc &= \begin{pmatrix} T_v & T_w \end{pmatrix},\quad
 \Tc^{-1}=\Tc^\top= \begin{pmatrix} T_v^\top \\ T_W^\top, \end{pmatrix}
\end{align}
where $v=\frac{1}{\sqrt{N}}\uno_N\in\Real^{N\times1}$ and $W\in\Real^{N\times(N-1)}$ are
as in Lemma \ref{lem:Somega}.
Consider the change of coordinates $\Et_t=\Tc^{-1} \Eb_t=\Tc^\top\Eb_t$.
In the new coordinates the error dynamics is
\begin{align}
   \Et_{t+1} =&  H_t(\o)\Et_t+\tilde{\hb}_t \label{eq:Etilde1} \\
H_t(\o)=&\begin{pmatrix} T_v^\top A_\Gc^{(\g)}(\omega)T_v &
            T_v^\top A_\Gc^{(\g)}(\omega) T_W \\
            T_W^\top A_\Gc^{(\g)}(\omega) T_v &
            T_W^\top A_\Gc^{(\g)}(\omega)T_W
           \end{pmatrix} \nonumber\\
&= \begin{pmatrix}
      A_C & T_v^\top\col_{i=1}^N(W_i\otimes A_i) \\
      (S_t^{(\g)}(\o)\otimes I_n)\mathcal{W}_A^\top T_v &
      (S_t^{(\g)}(\o)\otimes I_n)\mathcal{W}_A^\top T_W
     \end{pmatrix}  \nonumber\\
&=\begin{pmatrix}
     A_C & H_{12} \\ H_{21}^{(\g)}(\o) & H_{22}^{(\g)}(\o)
    \end{pmatrix} \label{eq:Htom}
\end{align}
with $\mathcal{W}_A^\top= \row_{i=1}^N(W_i^\top\otimes A_i)\ \in\Real^{nN\times n(N-1)}$, and
\begin{align}
 \tilde{\hb}_t &= \begin{pmatrix}
     (v^\top\otimes I_n)(M_t^{(\g)}(\o)\otimes I_n) \\ (W^\top\otimes I_n)(M_t^{(\g)}(\o)\otimes I_n)
              \end{pmatrix} \col_{i=1}^N(\hb_t^{(i)}) \nonumber \\
&=  \begin{pmatrix}
         T_v^\top  \hb_t \\
         (W^\top M^{(\g)}_t(\o)\otimes I_n) \hb_t
        \end{pmatrix} \label{eq:hbt}
\end{align}
where $W_i\in\Real^{1\times (N-1)}$ denotes the rows of $W=\col_{i=1}^N(W_i)$,
$\hb_t=\col_{i=1}^N(\hb_t^{(i)})$, $S_t^{(\g)}(\o)$ is defined by \eqref{eq:Stgamma}  and
we used Lemma \ref{lem:Mtgamma} to derive that
\begin{align}
  T_v^\top A_\Gc^{(\g)}(\omega) T_v &= \frac{1}{N}\sum_{i=1}^N A_i=A_C, \\
  T_v^\top A_\Gc^{(\g)}(\omega) T_W &= T_v^\top \col_{i=1}^N(W_i\otimes A_i),\\
 T_W^\top A_\Gc^{(\g)}(\omega)T_v &=
(W^\top M_t^{(\g)}(\o) \otimes I_n)\diag_{i=1}^N(A_i) T_v \nonumber\\
&= (S_t^{(\g)}(\o)\otimes I_n) \row_{i=1}^N (W_i^\top\otimes A_i) T_v.
\end{align}
Since $H_{21}^{(\g)}(\o)$, $H_{22}^{(\g)}(\o)$  are independent from the state and the noise
and $\E[\tilde{\hb}_t]=0$, by taking expectations in \eqref{eq:Etilde1},
\begin{align}
 &\E[\Et_{t+1}] 
  =\begin{pmatrix}
     A_C & T_v^\top \col_{i=1}^N(W_i\otimes A_i) \\
     (\bar{S}^{\g}\otimes I_n)\mathcal{W}_A^\top T_v & (\bar{S}^{\g}\otimes I_n)\mathcal{W}_A^\top T_W
    \end{pmatrix}\E[\Et_t], \label{eq:EEt}
\end{align}
where we have used Lemma \ref{lem:Mtgamma}, point 1, \eqref{eq:Sbar3} in Lemma \ref{lem:Sbar} and
the property $W^\top\bar{M}^{(\g)}=\bar{S}^{\g} W^\top$ that descends from \eqref{eq:Sbar2}
pre-multiplying by $W^\top$. We observe that $A_C=(I-K_\infty C)A$ is the dynamical matrix of the
centralized Kalman filter, which is Schur for Assumption \ref{ass:system}, and $\bar{S}$
is symmetric and Schur (see Lemma \ref{lem:Sbar}). Thus, $\lim_{\g\to\infty} \|\bar{S}^{\g}\|=0$
and there exists a sufficiently large $\g_0$ such that for all $\g>\g_0$ the dynamical matrix in
\eqref{eq:EEt} is Schur.

As for mean square boundedness, by invoking Lemma \ref{lem:random_stability}, iii), we just have
to prove that for a sufficiently large $\g$ the matrix $\E\left[H_t(\o)^{[2]}\right]$,
shown in Fig. \ref{fig:EH2tom}, is time-invariant and Schur.
Time invariance of $\E\left[H_t(\o)^{[2]}\right]$ is a consequence of the temporal independence of the blocks $H_{21}^{(\g)}(\o)$, $H_{22}^{(\g)}(\o)$.
From the structure of $H_{21}^{(\g)}$, $H_{22}^{(\g)}$ and
$\lim_{\g\to\infty}\|\E[{S_t^{(\g)}(\o)}^{[2]}]\|=0$ (Lemma \ref{lem:Mtgamma}, point 2-iv)
it follows that
\begin{align*}
&    \lim_{\g\to\infty}  \E\left[H_t(\o)^{[2]}\right] 
=    \begin{pmatrix}
    A_C^{[2]} & *
    \\ 0 &    0
    \end{pmatrix}
\end{align*}
and since $A_C^{[2]}$ is Schur, the desired property holds for $\g\to\infty$ and therefore
for a sufficiently large $\g$.

\subsection{Proof of Corollary \ref{cor:special} }\label{app:cor_special}

 We only need to prove \eqref{eq:Etilde1_special}.
In fact, from \eqref{eq:Etilde1_special} it follows that
 \begin{equation}
  \E[\Et_{t+1}]= \begin{pmatrix}
     A_C & 0 \\ 0 & \bar{S}^{\g} \otimes A_C
    \end{pmatrix}\E[\Et_t] \label{eq:EEtilde1_special}
 \end{equation}
and, since in the hypotheses both $A_C$ and $\bar{S}^{\g}$ are Schur, it follows
that the dynamical matrix is Schur. From \eqref{eq:Etilde1_special},
Lemma \ref{lem:Mtgamma}, 2) and Lemma \ref{lem:random_stability}, iii)
it follows also that $\Et_t$ is mean square bounded.
To prove \eqref{eq:Etilde1_special} we observe that
when $\forall i,j$ $A_i= A_j$,  then $A_i=A_C$ and
\begin{align}
 H_{12}=&  T_v^\top \col_{i=1}^N(W_i\otimes A_C)= T_v^\top (W\otimes A_C)=0 \nonumber,
\end{align}
and analogously $H_{21}^{(\g)}(\o)=0$ too. Moreover,
since $\sum_{i=1}^N W_i^\top W_i=I_{N-1}$,
\begin{align*}
 H_{22}^{(\g)}(\o)&= (S_t^{(\g)}(\o)\otimes I_n)\mathcal{W}_A^\top(W\otimes I_n)
 = S_t^{(\g)}(\o) \otimes A_C.
\end{align*}

\subsection{Proof of Lemma \ref{lem:tilde_h} }\label{app:lem_tilde_h}

Let $\widetilde{R}=\E[\row_i\col_j(h_t^{(i)}{h_t^{(j)}}^\top)]$ and $S_t^{(\g)}(\o)=\Pi_{h=0}^{\g-1}S_{t,h}(\o)$.
Since $\delta>\rho(\Lcb)$, $\lim_{\g\to\infty} S_t^{(\g)}(\o){S_t^{(\g)}(\o)}^\top$ exists.
We can apply Lebesgue's dominated convergence theorem to obtain
\begin{align*}
 \lim_{\g\to\infty} \Psi_{\hti} =&
   \lim_{\g\to\infty} \E\left[ \Tc^\top (M_t^{(\g)}(\o)\otimes I_n) \widetilde{R} (M_t^{(\g)}(\o)\otimes I_n)^\top \Tc \right]\nonumber\\
  =& \lim_{\g\to\infty} \E\Big[ \Tc^\top ((vv^\top+ WS_t^{(\g)}(\o)W^\top)\otimes I_n) \widetilde{R} \cdot \\
  & \cdot  (((vv^\top+ WS_t^{(\g)}(\o)^\top W^\top)\otimes I_n) \Tc \Big] \\
  =& \Tc^\top ((vv^\top\otimes I_n)\widetilde{R}(vv^\top\otimes I_n)\Tc 
  = \begin{pmatrix} T_v^\top \widetilde{R}T_v & 0 \\   0 & 0  \end{pmatrix}.
\end{align*}
The thesis follows by observing that
\begin{align}
 &T_v^\top \widetilde{R}T_v = \frac{1}{N}\sum_{i,j} \E\left[h_t^{(i)}{h_t^{(j)}}^\top\right] \nonumber\\
 &= N\left((I_n-K_\infty C)Q(I_n-K_\infty C)^\top+ P_\infty C^\top R^{-1}CP_\infty\right). \label{eq:Psih11}
\end{align}

\subsection{Proof of Theorem \ref{th:cov_err_i} }\label{app:th_cov_err_i}
The proof is obtained by showing that the covariance of the error in the transformed coordinates $\Pt_t= \E[\Et_t\Et_t^\top]$ satisfies
\begin{equation}\label{eq:Plimit}
 \lim_{t\to\infty}\lim_{\g\to\infty} \Pt_t= \begin{pmatrix} NP_\infty & 0 \\ 0 & 0 \end{pmatrix}
\end{equation}
that is equivalent to stating that the asymptotic (in time) limit (in $\g$) of the covariance of the
estimation error is identical across the $N$
agents and sums up to $N P_\infty$. Let $\Pt_0= T^\top \E[\Eb_0\Eb_0^\top]T$ be the initial
value of the covariance of the estimation error in the transformed coordinates.
By proceeding inductively, $\lim_{\g\to\infty} \Pt_{1}$ is equal to
\begin{align}
 &\lim_{\g\to\infty}
  \E\left[\begin{pmatrix} A_C & H_{12} \\ H_{21}^{(\g)}(\o) & H_{22}^{(\g)}(\o)
    \end{pmatrix} \Pt_{0}\begin{pmatrix} A_C^\top & {H^{21}}^\top_\g(\o) \\ {H_{12}}^\top & H_{22}^{(\g)}(\o)
    \end{pmatrix}\right] \nonumber \\
    &\qquad \qquad\qquad \qquad\qquad \qquad\qquad \qquad\qquad \qquad + \lim_{\g\to\infty} \Psi_{\hti} \nonumber \\
&=
  \begin{pmatrix} A_C & H_{12} \\ 0 & 0
    \end{pmatrix} \Pt_{0}\begin{pmatrix} A_C^\top & 0 \\{H_{12}}^\top & 0
    \end{pmatrix} + \lim_{\g\to\infty} \Psi_{\hti} \nonumber
    \end{align}
    \begin{small}
\begin{align}
= \begin{pmatrix}
        A_C (\Pt_0)_{1,1} A_C^\top + H_{12}(\Pt_0)_{1,2}+(\Pt_0)_{1,2}^\top {H_{12}}^\top
        +(\Psi_{\hti})_{1,1} & 0\\ 0 & 0
       \end{pmatrix}, \nonumber
\end{align}
\end{small}
where $ (\Psi_{\hti})_{1,1}$ is given by \eqref{eq:Psih11}, namely
\begin{equation}
 (\Psi_{\hti})_{1,1}= N(I_n-K_\infty C)Q(I_n-K_\infty C)^\top+ N P_\infty C^\top R^{-1}CP_\infty. \nonumber
\end{equation}
Therefore, only the first $n\times n$ block of $\lim_{\g\to\infty} \Pt_{1}$ is not zero,  and
this property continues to hold for all the matrices $\lim_{\g\to\infty} \Pt_{t}$.
The block $\frac{1}{N}\Pt^{1,1}_t$ evolves as
\begin{align}
 \frac{1}{N}\Pt^{1,1}_{t+1}&= (I_n-KC)(A \frac{1}{N} \Pt^{1,1}_{t} A^\top+Q)(I_n-KC)^\top \nonumber\\
 &\quad +  P_\infty C^\top R^{-1}CP_\infty.
\end{align}
Since $P_\infty$ satisfies the algebraic Riccati equation \eqref{eq:ric_CKF}
the difference $\Delta_t=P_\infty- \frac{1}{N}\Pt^{1,1}_{t}$ evolves as
$\Delta_{t+1}= A_C \Delta_t A_C^\top$,
which is an asymptotically stable matrix function because $A_C$ is Schur.
Consequently, $\lim_{t\to\infty}\frac{1}{N}\Pt^{1,1}_{t}= P_\infty$.
\hfill$\Box$

\section{A result on the stability of random matrices}\label{app:random_matrices}

In a probability space $(\Omega,\Fc,P)$ the linear space of $\Fc$-measurable vectors $x: \Omega\to\Real^n$
with finite second moment, endowed with the inner product $[x_1,x_2]_{\Lc_2}=\E[x_1^\top x_2]$
is a Hilbert space denoted $\Lc_2^n=L_2((\Omega,\Fc,P);\Real^n)$. If $x\in\Lc_2^n$
then $\E[\|x\|^2]<\infty$. In particular, when $\E[x]=0$ then $\|x\|^2_{\Lc_2^n}$ is the variance of the random vector $x$. A random square matrix $A(\o):\ \Omega\to\Real^{n\times n}$ is a $\Fc$-measurable function from $\Omega$ to the set of square matrices of size $n$.

\begin{Lemma}\label{lem:random_stability}
Let $\{A(t,\o)\}$, $t=0,1,\ldots$, be a sequence of independent, $\Fc$-measurable and identically distributed
matrix functions $A(t,\o):\ \Omega\to\Real^{n\times n}$.
 Given the system
 \begin{equation}\label{eq:generic_sys}
  x(t+1)= A(t,\o)x(t)+ f(t,\o),
 \end{equation}
where $x(0)\in\Lc^n_2$, $f(t,\o)$ is a sequence of mutually independent vectors in $\Lc^n_2$, with $\E[f(t,\o)]=0$ and $f(t,\o)$ independent from $x_0$ for all $t$, $A(t,\o)$ independent from $x(0)$ and $f(t,\o)$, then
\begin{enumerate}
 \item [i)] if $\rho(\E[A])<1$ then $\E[x(t)]\to 0$ exponentially.
 \item [ii)] if $\E[\|A\|^2]<1$ then $\|x(t)\|_{\Lc_2}$ is uniformly bounded for all $t$.
 \item [iii)] if $\rho(\E[A^{[2]}])<1 $ then $\|x(t)\|_{\Lc_2}$ is uniformly bounded for all $t$.
\end{enumerate}
\end{Lemma}
\textit{Proof.}
i): At each $t$ $A(t,\o)$ and $x(t)$ are independent,
\begin{equation}
 \E[x(t+1)]= \E[Ax(t)]= \E[A]\E[x(t)]
\end{equation}
and since $\E[A]$ is Schur, $\E[x(t)]\to0$.

ii):
\begin{align}
 \|x(t+1)\|_{\Lc_2} &= \|  A(t,\o)x(t)+ f(t,\o)\|_{\Lc_2} \nonumber \\
 &\quad \leq \E[\|A\|^2]^{\frac{1}{2}} \|x(t)\|_{\Lc_2}+ \|f(t,\o)\|_{\Lc_2}.
\end{align}

iii): for any conformable matrices $A$, $B$, $C$ it holds that $\st(ABC)=(C^\top\otimes A)\cdot\st(B)$. Let $P(t)=\E[x(t)x(t)^\top]$. \eqref{eq:generic_sys} implies that
\begin{align}
P(t+1)&= \E\left[A(t,\o)P(t)A(t,\o)^\top\right]+ \E[f(t,\o)f(t,\o)^\top] \\
\st(P(t+1)) &= \E\left[A(t,\o)^{[2]} \right] \st(P(t)) \nonumber \\
&\quad\qquad \qquad + \st( \E[f(t,\o)f(t,\o)^\top]). \label{eq:stack_P}
\end{align}
It follows that if $\E[A(t,\o)^{[2]}]$ is Schur then $\st(P(t))$ is uniformly bounded
in time and the same holds for $P(t)$. Since $\|x(t)\|^2_{\Lc_2}=\tr(P(t))$ the thesis follows.

\hfill $\Box$

\bibliographystyle{plain}
\bibliography{biblio.bib}

\end{document}